\documentclass[12pt,a4paper]{article}
\usepackage{amsmath}
\usepackage{amsfonts}
\usepackage{graphicx}
\usepackage{cite}

\DeclareMathOperator{\Li}{Li}

\def\({\left(}
\def\){\right)}

\newcommand{\ket}[1]{|#1\rangle}
\newcommand{\bra}[1]{\langle #1 |}

\newcommand{\nn}{\nonumber}
\newcommand{\Eqn}[1]{&\hspace{-0.5em}#1\hspace{-0.5em}&}
\renewcommand{\[}{\begin{equation}}
\renewcommand{\]}{\end{equation}}
\newcommand{\eqb}{\begin{eqnarray}}
\newcommand{\eqe}{\end{eqnarray}}

\newcommand{\bbC}{{\mathbb C}}
\newcommand{\bbR}{{\mathbb R}}
\newcommand{\bbZ}{{\mathbb Z}}

\newcommand{\grp}[1]{\mathrm{#1}}

\newcommand{\bm}{{\bar{m}}}
\newcommand{\bw}{{\bar{w}}}
\newcommand{\bz}{{\bar{z}}}

\newcommand{\cK}{{\cal K}}

\newcommand{\halpha}{{\hat{\alpha}}}

\newcommand{\tR}{{\tilde{R}}}
\newcommand{\tY}{{\tilde{Y}}}

\newcommand{\dilog}{\operatorname{Li_2}}
\newcommand{\tepsilon}{{\tilde{\epsilon}}}

\def\varep{\varepsilon}

\def\bra{\langle}
\def\ket{\rangle}

\def\cdots {\cdot\cdot\cdot}
\def\pint#1 {- \!\!\!\!\!\!\!\! \,\int_{#1}}
\def\semiket#1  { \, #1 \, \rangle \, }
\def\del        {  \partial  }

\def\half       {\frac{1}{2}}
\def\abs#1      {  \, \vert #1 \vert \,   }
\def\Im#1    { \, {\rm Im } \, #1  }
\def\Re#1    { \, {\rm Re}  \, #1  }

\def\trans#1  { {}^{t} \! \vec{#1} }
\def\binom#1#2 { \vecii{ {}_{#1} }{\raisebox{.5ex}{$ {}^{#2} $}} }
\def\sqbinom#1#2 { \Bigl(\begin{array}{c} {}_{#1}
                       \\ \raisebox{.5ex}{${}^{#2}$}
                         \end{array}\Bigr)^2  }

\def\zbar   {\bar{z}}

\def\r12    {\frac{r_1}{r_2}}

\def\vecii#1#2     {{ #1 \choose #2 }  }
\def\veciii#1#2#3  {\left(\begin{array}{c}#1\\#2\\#3\end{array}\right)}
\def\matrixii#1#2#3#4            {\Bigl( \begin{array}{cc}#1&#2\\#3&#4
                                     \end{array} \Bigr) }
\def\matrixiii#1#2#3#4#5#6#7#8#9 {\left(\begin{array}{ccc}#1&#2&#3\\
                                  #4&#5&#6\\#7&#8&#9\end{array}\right)}

\renewcommand{\thesection}
  {\arabic{section}.\hspace{-.5em}}
\renewcommand{\thesubsection}
  {\arabic{section}.\arabic{subsection}.\hspace{-.5em}}
\renewcommand{\thesubsubsection}
  {\arabic{section}.\arabic{subsection}.\arabic{subsubsection}.\hspace
                                                               {-.5em}}

\makeatletter
\renewcommand\section{
  \@startsection{section}{3}{\z@}%
  {-3.25ex\@plus -1ex \@minus -.2ex}%
  {1.5ex \@plus .2ex}%
  {\normalfont\normalsize\bfseries\mathversion{bold}}}
\renewcommand\subsection{
  \@startsection{subsection}{3}{\z@}%
  {-3.25ex\@plus -1ex \@minus -.2ex}%
  {1.5ex \@plus .2ex}%
  {\normalfont\normalsize\bfseries\mathversion{bold}}}
\renewcommand\subsubsection{
  \@startsection{subsubsection}{3}{\z@}%
  {-3.25ex\@plus -1ex \@minus -.2ex}%
  {1.5ex \@plus .2ex}%
  {\normalfont\normalsize\itshape}}
\makeatother

\makeatletter \@addtoreset{equation}{section} \makeatother
\renewcommand{\theequation}{\arabic{section}.\arabic{equation}}

\makeatletter
\renewcommand{\appendix}{
\renewcommand{\thesection}{Appendix \Alph{section}.\hspace{-.5em}}
\renewcommand{\thesubsection}
  {\Alph{section}.\arabic{subsection}.\hspace{-.5em}}
\renewcommand{\thesubsubsection}
  {\Alph{section}.\arabic{subsection}.\arabic{subsubsection}.\hspace
                                                             {-.5em}}
\@addtoreset{equation}{subsection}
\renewcommand{\theequation}{\Alph{section}.\arabic{equation}}
\setcounter{section}{0}}
\makeatother

\begin{document}
%
\def\papertitlepage{\baselineskip 3.5ex \thispagestyle{empty}}
\def\preprinumber#1#2#3#4{\hfill
\begin{minipage}{1.2in}
#1 \par\noindent #2 \par\noindent #3 \par\noindent #4
\end{minipage}}
\renewcommand{\thefootnote}{\fnsymbol{footnote}}
\newcounter{aff}
\renewcommand{\theaff}{\fnsymbol{aff}}
\newcommand{\affiliation}[1]{
  \setcounter{aff}{#1} $\rule{0em}{1.2ex}^\theaff\hspace{-.4em}$}
%
%
\papertitlepage
\setcounter{page}{0}
\preprinumber{YITP-10-38}{TIT/HEP-603}{UTHEP-607}{arXiv:1005.4487}
\vskip 2ex
~
\vfill
\begin{center}
{\large\bf\mathversion{bold}
Six-point gluon scattering amplitudes\\
from $\bbZ_4$-symmetric integrable model}
\end{center}
\vfill
\baselineskip=3.5ex
\begin{center}
  Yasuyuki Hatsuda\footnote[1]{\tt hatsuda@yukawa.kyoto-u.ac.jp}, 
  Katsushi Ito\footnote[2]{\tt ito@th.phys.titech.ac.jp},
  Kazuhiro Sakai\footnote[3]{\tt sakai@phys-h.keio.ac.jp} and
  Yuji Satoh\footnote[4]{\tt ysatoh@het.ph.tsukuba.ac.jp}\\

\vskip 2ex
  \affiliation{1}
  {\it Yukawa Institute for Theoretical Physics, Kyoto University}\\
  {\it Kyoto 606-8502, Japan}\\

\vskip 1ex
  \affiliation{2}
  {\it Department of Physics, Tokyo Institute of Technology}\\
  {\it Tokyo 152-8551, Japan}\\

\vskip 1ex
  \affiliation{3}
  {\it Research and Education Center for Natural Sciences}\\
  {\it and Hiyoshi Department of Physics, Keio University}\\
  {\it Yokohama 223-8521, Japan}\\

\vskip 1ex
  \affiliation{4}
  {\it Institute of Physics, University of Tsukuba}\\
  {\it Ibaraki 305-8571, Japan}

\end{center}
\vfill
%
\baselineskip=3.5ex
\begin{center} {\bf Abstract} \end{center}

We study six-point gluon scattering amplitudes in
${\cal N} = 4$ super Yang--Mills theory at strong coupling
by  investigating the thermodynamic Bethe ansatz equations of
the underlying $\bbZ_4$-symmetric integrable model both
analytically and numerically. 
By the conformal field theory (CFT) perturbation,
we compute  the free energy part of the remainder function
with generic chemical potential 
near the CFT/small mass limit. Combining this with 
the expansion of the Y-functions, we obtain the remainder function
near the small mass limit 
up to a function of the chemical potential, which can be evaluated 
numerically.
We also find the leading corrections to the remainder function
near the large mass limit.  We confirm that these results are
in good agreement with numerical computations.

\vskip 2ex
\vspace*{\fill}
\noindent
May 2010
\setcounter{page}{0}
\newpage
\renewcommand{\thefootnote}{\arabic{footnote}}
\setcounter{footnote}{0}
\setcounter{section}{0}
\baselineskip = 3.5ex
\pagestyle{plain}
%

\section{Introduction}

Recently there has been much progress in computing the area of
minimal surfaces in AdS space, whose boundary is made of light-like
segments \cite{Alday:2007hr,Alday:2009yn,Alday:2009dv,Burrington:2009bh,
Alday:2010vh,Hatsuda:2010cc,Yang:2010az}.
The surface area corresponds to the expectation value of the 
Wilson loop along the same contour and is dual to
the gluon scattering amplitude
in planar ${\cal N}=4$ super Yang--Mills theory
at strong coupling.
The minimal surface area has deviation from the Bern--Dixon--Smirnov
(BDS) conjecture \cite{Bern:2005iz}
by an amount of the remainder function,
which is shown to exist for the $n(\geq\! 6)$-point
amplitudes \cite{Alday:2007he,Dr2}.
Determination of the remainder function
is a key step toward establishing the correct analytic formula
of the gluon scattering amplitude.

In \cite{Alday:2009dv}, the equations for determining the minimal
surface in AdS$_5$ space are shown to be the $\grp{SU}(4)$ Hitchin
equations with certain constraints and boundary conditions.
The area is characterized by the Stokes data of the asymptotic 
solutions of the associated linear system.
The Stokes data obey certain functional equations
and the minimal surface area 
is evaluated by solving the integral equations associated with them.
For the 6-point gluon scattering amplitudes, 
the functional equations and the integral equations turn out to be
the Y-system \cite{Zamolodchikov:1991et} and
the thermodynamic Bethe ansatz (TBA)
equations \cite{Zamolodchikov:1989cf}
of the $\bbZ_4$-symmetric (or $A_3$)  integrable model 
\cite{Koberle:1979sg,Tsvelick88,Fateev:1990bf}, respectively.
The minimal surface area is evaluated by the free energy of the 
model.

This solution has been generalized to the minimal surface with an
$n$-sided light-like polygonal boundary
in AdS$_5$ \cite{Alday:2010vh} and in AdS$_3$
\cite{Alday:2010vh,Hatsuda:2010cc}.
In particular, Alday, Maldacena, Sever and Vieira proposed the 
Y-system and the TBA equations for the $n$-sided polygonal solution
and expressed the area in terms of the TBA system.
In \cite{Hatsuda:2010cc}, the present authors noticed that
the TBA system is that of the homogeneous sine-Gordon model
\cite{FernandezPousaEtAl}.

The free energy for a two-dimensional integrable model with purely 
elastic S-matrix is obtained by solving the TBA system.
It is very difficult to solve the TBA system exactly but one can 
investigate its solutions
in two limits.
In the UV (or high temperature or small mass) limit, it is described by
a certain conformal field theory (CFT).
In the IR (or low temperature or large mass) 
limit, on the other hand,
it becomes a system of a free massive theory.
In the UV limit, the free energy turns out to be the central charge
of the CFT.
In the case of the homogeneous sine-Gordon model,
the relevant CFTs are
the generalized parafermion CFTs \cite{parafermions}. 
A particularly interesting feature of the TBA  system is that it
contains chemical potential which arises from the monodromy of the
asymptotic solution of the linear system.
Its solution with chemical potential
near the CFT point has not been well studied.

While analysis in the above described limits reveals
several principal characters of the amplitude,
evaluation of the remainder function apart from the limits
is also of great significance
in studying how the amplitude depends on the gluon momenta.
The purpose of this paper is to analyze
such momenta dependence of the remainder function.
To do this, we investigate the TBA system
perturbatively both near the UV and IR limits.
In the present work we will focus on the 6-point amplitude for
simplicity, where the relevant CFT is the $ \bbZ_4$-parafermion CFT.
The solution of the TBA system without chemical potential
around the CFT point 
was studied by Klassen and Melzer \cite{KlaMe}.
In this paper we will investigate the perturbative solution around the
CFT point including chemical potential.
We also study the TBA system near the IR limit and present 
the first correction to the remainder function.

Our results provide an analytic form of the remainder function 
for the 6-point scattering amplitude at strong coupling away from the
CFT and the IR point.
This paper also demonstrates that 
the unexpected connection between the four-dimensional super 
Yang--Mills theory and the two-dimensional integrable models discovered
in \cite{Alday:2009dv,Hatsuda:2010cc}
enables one to compute the amplitude at strong coupling
in two-dimensional approaches. 
From a point of view of the study of integrable models,
our discussion provides a concrete example of an analysis
of TBA systems with chemical potential.

This paper is organized as follows.
In sect.~2, we review the construction of the Hitchin equations of the
minimal surface with a 6-sided polygonal boundary 
in AdS$_5$ and the related Y-system and TBA equations.
In sect.~3, we study the CFT limit of the TBA system and evaluate the 
free energy perturbatively.
In sect.~4, we study the remainder function around the CFT limit.
In sect.~5, we examine the large mass limit and obtain the correction to
the remainder function. We conclude with a discussion in sect.~6.

\section{Review of TBA system for six-point amplitudes}

In this section we review the TBA system for the six-point
gluon scattering amplitudes.
We basically follow the references \cite{Alday:2009dv,Alday:2010vh}.

\subsection{Classical string solutions with a null polygonal boundary}

Alday and Maldacena proposed a method of computing
gluon scattering amplitudes in ${\cal N}=4$ super Yang--Mills
using AdS/CFT correspondence \cite{Alday:2007hr}.
According to their proposal, scalar magnitude of
$n$-gluon MHV scattering amplitudes
can be evaluated in the strong coupling limit
by computing the area of corresponding classical open string solutions.
The string solutions are minimal surfaces whose boundary
is a polygon located on the boundary of $AdS_5$.
The polygon consists of $n$ null edges
given by the $n$ momenta of incoming gluons.

To be concrete, classical string solutions under consideration
are realized as a map from their world-sheet to $AdS_5$.
We consider Euclidean world-sheet parametrized by $z,\bz$.
In terms of the global coordinates,
$AdS_5$ is expressed as a hypersurface
\eqb
\vec{X}\cdot\vec{X}\equiv
-(X^{-1})^2-(X^0)^2+(X^1)^2+(X^2)^2+(X^3)^2+(X^4)^2=-1
\eqe
in $\bbR^{2,4}$.
Classical string solutions $\vec{X}(z,\bz)$
satisfy the equations of motion
\eqb\label{EOM}
\partial\bar\partial\vec{X}
- (\partial \vec{X} \cdot \bar\partial \vec{X}) \vec{X}=0
\eqe
and the Virasoro constraints
\eqb\label{Virasoro}
\partial\vec{X}\cdot\partial\vec{X}=
\bar\partial\vec{X}\cdot\bar\partial\vec{X}=0.
\eqe

The boundary condition is expressed
in terms of the Poincar\'e coordinates $(r,x^\mu)$ given by
\eqb
X^\mu \Eqn{=} \frac{x^\mu}{r},\quad \mu=0,1,2,3,\\
X^{-1}+X^4\Eqn{=}\frac{1}{r}\qquad X^{-1}-X^4=\frac{r^2+x^\mu x_\mu}{r}.
\eqe
The solutions which correspond to $n$-gluon amplitudes
end on an $n$-sided polygon at the AdS boundary $r=0$.
The vertices $x_1,\ldots,x_n$ of the polygon are separated
by the gluon momenta $k_1,\dots, k_n$ as
\eqb
x_j^\mu-x_{j+1}^\mu=k_j^\mu.
\eqe

This type of boundary condition is neatly characterized
in terms of the generalized sinh-Gordon
potential \cite{Pohlmeyer:1975nb}.
Let us introduce the following notation
\eqb
e^{\alpha(z,\bz)}\Eqn{=}\partial\vec{X}\cdot\bar\partial\vec{X},\\
P(z)\Eqn{=}\partial^2\vec{X}\cdot\partial^2\vec{X},\qquad
\bar P(z)=\bar\partial^2\vec{X}\cdot\bar\partial^2\vec{X},\\
\halpha(z,\bz)\Eqn{=}\alpha(z,\bz)-\frac{1}{4}\log P(z)\bar{P}(\bz),
\eqe
where $\alpha,\halpha$ are real and $P(z)$ is shown to be
analytic in $z$.
For the simplest four-cusp solution,
$P(z)=1$ and $\halpha=0$ on the whole $z$-plane.
For $n$-cusp solutions,
$P(z)$ is a polynomial of degree $n-4$
and $\halpha\to 0$ for $|z|\to\infty$.
The latter condition reflects the fact that
$n$-cusp solutions have the same asymptotics
with the four-cusp solution around each cusp.

In this paper we will concentrate on the case of $n=6$,
where $P(z)$ is quadratic.
One can choose a gauge
\eqb\label{Pofz}
P(z)=z^2-U,\qquad U\in\bbC
\eqe
by a suitable redefinition of the world-sheet coordinate.

\subsection{Hitchin system}

The equations of motion (\ref{EOM}) and the Virasoro constraints
(\ref{Virasoro}) can be rephrased as
$\grp{SU}(4)$ Hitchin equations \cite{Alday:2009dv}.
To see this, let us consider a moving-frame basis spanned by
$\vec{q}_0=\vec{X},\vec{q}_4=e^{-\alpha/2}\partial\vec{X},
\vec{q}_5=e^{-\alpha/2}\bar\partial\vec{X}$
and the other three complementary orthonormal vectors
$\vec{q}_I,\ I=1,2,3$.
The evolution of the basis
$q=(\vec{q_0},\ldots,\vec{q_5})$
is described by a set of linear differential equations
\eqb\label{evolutioneq}
\partial q = -{\cal A}_z q,\qquad
\bar\partial q = -{\cal A}_\bz q.
\eqe
One can decompose the connection into two
parts ${\cal A}=A+\Phi$,
in such a way that $A$ rotates
$(\vec{q}_0,\vec{q}_1,\vec{q}_2,\vec{q}_3)$
and $(\vec{q}_4,\vec{q}_5)$ separately among
themselves while $\Phi$ mixes them.
The flatness condition of (\ref{evolutioneq})
then takes the form of the Hitchin equations
\eqb
\label{Hitchineq1}
D_z\Phi_\bz=0,\quad D_\bz\Phi_z=0,\\
\label{Hitchineq2}
{}[D_z,D_\bz]+[\Phi_z,\Phi_\bz]=0,
\eqe
where
$D_z=\partial+[A_z,\ \ ],\ D_\bz=\bar\partial+[A_\bz,\ \ ]$.
These equations are equivalent to the equations of motion (\ref{EOM})
and the Virasoro constraints (\ref{Virasoro}).
One can write down the same equations in the spinor basis,
where $A_z,A_\bz,\Phi_z,\Phi_\bz$
now represent $4\times 4$ matrices.
We do not need their explicit form \cite{Alday:2009dv} here,
but an important fact is that these
$A$ and $\Phi$ corresponding to string solutions
satisfy additional constraints
\eqb
CA^{\rm T}C^{-1}=-A,\qquad C\Phi^{\rm T}C^{-1}=i\Phi,
\eqe
with a certain constant matrix $C$.
This leads to a $\bbZ_4$ automorphism in the present Hitchin system.
This $\bbZ_4$ automorphism is not inherent
in general Hitchin systems but is peculiar to the one
describing the classical strings in $AdS_5$.
The system also exhibits $\bbZ_2$ automorphism that
corresponds to the reality of the string solutions. 

As argued above, $n$-cusp solutions are characterized
by the condition that $\halpha\to 0$ for $|z|\to\infty$.
This is equivalent to the statement that one can always
diagonalize $\Phi$ and $A$ at large $|z|$
by a suitable gauge transformation
into the following form
\eqb
h^{-1}\Phi_z h\Eqn{\to}\frac{1}{\sqrt{2}}
\left(\begin{array}{cccc}
P(z)^{1/4}&&&\\
&-iP(z)^{1/4}&&\\
&&-P(z)^{1/4}&\\
&&&iP(z)^{1/4}
\end{array}\right),\\
h^{-1}A_z h + h^{-1}\partial h
\Eqn{\to}\frac{m}{z}\left(\begin{array}{cc}
\sigma_3&0\\
0&\sigma_3\end{array}
\right),\qquad
h^{-1}A_\bz h + h^{-1}\bar\partial h
\to - \frac{\bar m}{\bz}\left(\begin{array}{cc}
\sigma_3&0\\
0&\sigma_3\end{array}
\right),\nn\\
\eqe
where $m$ and $\bar{m}$ are constants.

\subsection{Auxiliary linear problem and small solutions}

The system under consideration is classically integrable.
This means that the connections appeared in the linear equations
can be promoted to a set of one-parameter family of flat connections.
That is to say,
one can write down an auxiliary linear problem
\eqb\label{auxlineq}
\nabla_z^\zeta q(z,\bz;\zeta)=0,\qquad
\nabla_\bz^\zeta q(z,\bz;\zeta)=0,
\eqe
with 
\eqb
\nabla_z^\zeta =D_z+\zeta^{-1}\Phi_z,\qquad
\nabla_\bz^\zeta =D_\bz+\zeta\Phi_\bz,
\eqe
where the Hitchin equations (\ref{Hitchineq1}), (\ref{Hitchineq2})
are obtained as the compatibility condition
\eqb
[\nabla_z^\zeta,\nabla_\bz^\zeta]=0.
\eqe

One can estimate the asymptotic form of the solutions
$q(z,\bz;\zeta)$ for large $|z|$.
There are four independent solutions, whose asymptotic forms
are respectively given by
\eqb
&\vec{e}_1
z^m    \bz^{-\bar m}e^{ \frac{1}{\sqrt{2}}(\zeta^{-1}w+\zeta\bw)},\qquad
 \vec{e}_2
z^{-m} \bz^{ \bar m}e^{-\frac{i}{\sqrt{2}}(\zeta^{-1}w-\zeta\bw)},&\nn\\
&\vec{e}_3
z^m    \bz^{-\bar m}e^{-\frac{1}{\sqrt{2}}(\zeta^{-1}w+\zeta\bw)},\qquad
 \vec{e}_4
z^{-m} \bz^{ \bar m}e^{ \frac{i}{\sqrt{2}}(\zeta^{-1}w-\zeta\bw)},&
\eqe
where $w=\int^z P(z)^{1/4}dz\sim z^{n/4}$
and $\vec{e}_i$ are constant vectors.
One can determine which of the four solutions
shows the fastest decay for $|z|\to\infty$
in each sector
\eqb
W_k: \frac{\pi(2k-3)}{n}+\frac{4}{n}\arg\zeta <
\arg z < \frac{\pi(2k-1)}{n}+\frac{4}{n}\arg\zeta.
\eqe
We call such solution the small solution $s_k(z,\bz;\zeta)$
in the sector $W_k$.
Small solutions form a set of redundant basis of
the solutions to the equations (\ref{auxlineq}).
We normalize the solutions so that
\eqb\label{s_normcond}
\langle s_j,s_{j+1},s_{j+2},s_{j+3}\rangle=1,
\eqe
where 
$\langle s_i,s_j,s_k,s_l\rangle \equiv\det(s_i s_j s_k s_l)$.
As the differential operators are regular everywhere in $|z|<\infty$,
the sectors $W_j$ and $W_{j+n}$ are identified. Hence
\eqb
s_{j+n}\propto s_{j}.
\eqe
One can determine the proportionality coefficient
case by case whether $n=2k-1$, $n=4k-2$
or $n=4k$ for $k\in\bbZ_{>0}$.
In the case of $n=6$,
\eqb\label{s_identity}
s_{j+6}=\mu^{(-1)^j} s_j
\eqe
where
\eqb
\mu=\pm e^{2\pi i(m+\bm)}.
\eqe
Note that $\mu$ is a pure phase factor
for solutions in the usual $(3,1)$ signature.
For later use, let us also introduce the notation
\eqb
\mu=e^{i\phi}
\eqe
with $\phi$ being a real parameter.

\subsection{Y-system}

Making use of integrability,
one can compute conserved quantities
without knowing the explicit form of solutions.
Instead, the fundamental building blocks
are the Stokes data $\langle s_i,s_j,s_k,s_l\rangle(\zeta)$.
Let us first recall some important identities concerning them.
In addition to the normalization condition (\ref{s_normcond}),
there hold the following identities
\eqb
\label{StokesZ4sym1}
\langle s_k,s_{k+1},s_j,s_{j+1}\rangle(\zeta)
\Eqn{=}\langle s_{k-1},s_k,s_{j-1},s_j\rangle(i\zeta),\\
\label{StokesZ4sym2}
\langle s_j,s_k,s_{k+1},s_{k+2}\rangle(\zeta)
\Eqn{=}\langle s_j,s_{j-1},s_{j-2},s_k\rangle(i\zeta),
\eqe
which follow from the $\bbZ_4$ automorphism.

In \cite{Alday:2010vh} Alday, Maldacena, Sever and Vieira
formulated the T-system and the Y-system associated
to the general $n$-cusp solutions in $AdS_5$.
There, some particular Stokes data are chosen
as T-functions and subsequently Y-functions
are defined as ratios of the form $Y=TT/TT$.
Similar formulation of T- and Y-system
was obtained from the spectral theory of 
ordinary differential equations \cite{Dorey:2007zx}.
By restricting ourselves to the $n=6$ case,
which is the simplest nontrivial case in their formulation,
general formulas partly get simplified
due to the relation (\ref{s_normcond}).
Y-functions are then defined by\footnote{
Y-functions introduced here are identified with those
in \cite{Alday:2010vh} as
$
Y_1(\theta)=\mu^{-1}[Y_{1,1}^{\rm AMSV}(ie^\theta)]^{-1}$, \ $
Y_2(\theta)=[Y_{2,1}^{\rm AMSV}(ie^\theta)]^{-1}$, \ $ 
Y_3(\theta)=\mu [Y_{3,1}^{\rm AMSV}(ie^\theta)]^{-1}.
$
}
\eqb
\label{defY1}
Y_1(\theta)\Eqn{=}
-\langle s_2,s_3,s_5,s_6\rangle(e^{\theta}),\\
\label{defY2}
Y_2(\theta)\Eqn{=}
\langle s_1,s_2,s_3,s_5\rangle
\langle s_2,s_4,s_5,s_6\rangle(e^{\theta+\pi i/4}),\\
\label{defY3}
Y_3(\theta)\Eqn{=}Y_1(\theta).
\eqe
These Y-functions are not entirely independent, but satisfy
functional relations.
By using Hirota bilinear identities (or Pl\"ucker relations)
among determinants and (\ref{s_identity})
(\ref{StokesZ4sym1}), (\ref{StokesZ4sym2}),
one can show that
\eqb
\label{Ysystem1}
Y_1\left(\theta+\frac{\pi i}{4}\right)
Y_1\left(\theta-\frac{\pi i}{4}\right)
\Eqn{=}1+Y_2(\theta),\\
\label{Ysystem2}
Y_2\left(\theta+\frac{\pi i}{4}\right)
Y_2\left(\theta-\frac{\pi i}{4}\right)
\Eqn{=}\bigl(1+\mu Y_1(\theta)\bigr)
  \bigl(1+\mu^{-1} Y_1(\theta)\bigr).
\eqe
Another important property of the Y-functions
is the periodicity
\eqb\label{Yperiodicity}
Y_a\left(\theta+\frac{3\pi i}{2}\right)\Eqn{=}Y_a(\theta),
\eqe
which also follows from
(\ref{s_identity}), (\ref{StokesZ4sym1}), (\ref{StokesZ4sym2}).
They also satisfy
\eqb\label{Yreality}
\overline{Y_a(\theta)}\Eqn{=}Y_a(-\bar\theta),
\eqe
which follows from the reality of string solutions.

In fact, the Y-system obtained here is identified
with that \cite{Zamolodchikov:1991et}
of $\bbZ_N$-symmetric integrable models
\cite{Koberle:1979sg,Tsvelick88,Fateev:1990bf} with $N=4$.
$Y_a$ correspond to the fundamental representations
labeled by $a=1,2,3$ of the $A_3$ Lie algebra.
The present Y-system
corresponds to the model with a chemical potential $\mu$ turned on.
Note that the periodicity (\ref{Yperiodicity})
holds in the presence of $\mu$.

The cross-ratios of gluon momenta are given by
special values of the Y-functions,
\eqb\label{Yspecial}
b_k=Y_1\left(\frac{(k-1)\pi i}{2}\right),\qquad
U_k=1+Y_2\left(\frac{(2k+1)\pi i}{4}\right),
\eqe
for $k=1,2,3$,
where
\eqb\label{crossratios}
U_1=b_2b_3=\frac{x_{14}^2x_{36}^2}{x_{13}^2x_{46}^2},\qquad
U_2=b_3b_1=\frac{x_{25}^2x_{14}^2}{x_{24}^2x_{15}^2},\qquad
U_3=b_1b_2=\frac{x_{36}^2x_{25}^2}{x_{35}^2x_{26}^2}.
\eqe
From the Y-system relations
(\ref{Ysystem1}), (\ref{Ysystem2}) one can verify that
$b_j$'s are not entirely independent but obey the constraint
\eqb\label{b_constraint}
b_1b_2b_3=b_1+b_2+b_3+\mu+\mu^{-1}.
\eqe
%

\subsection{TBA equations}

The functional relations described above
constrain the form of Y-functions mostly, but not entirely.
To fully determine the Y-functions,
we need an additional information
such as the asymptotic behavior and the singularity structure.
The asymptotic behavior of the Y-functions can be evaluated by
the WKB analysis \cite{Gaiotto:2009hg}.
It can be shown that
the Y-functions defined by (\ref{defY1})--(\ref{defY3})
exhibit the following asymptotics
\eqb
\log Y_1(\theta)\to |Z|e^{\pm(\theta-i\varphi)},\qquad
\log Y_2(\theta)\to
\sqrt{2}|Z|e^{\pm(\theta-i\varphi)}\nn\\[1ex]
\mbox{for}\quad{\rm Re\,}\theta\to\pm\infty,
\qquad
\varphi-\frac{\pi}{4}<{\rm Im\,}\theta<\varphi+\frac{\pi}{4},
\eqe
where $Z$ is related to the moduli parameter $U$ in (\ref{Pofz}) as
\eqb\label{ZUrel}
Z\equiv|Z|e^{i\varphi}
=U^\frac{3}{4}\int_{-1}^1(1-t^2)^\frac{1}{4}dt
=\frac{\sqrt{\pi}\Gamma(\frac{1}{4})}{3\Gamma(\frac{3}{4})}
  U^\frac{3}{4}.
\eqe
As for the singularity structure,
one can take Y-functions regular everywhere except
at ${\rm Re\,}\theta=\pm \infty$.
This is consistent with
the functional relations (\ref{Ysystem1}), (\ref{Ysystem2}).
One could also consider Y-functions with singularities
at finite $\theta$. These correspond to excited states.
Since we are interested in minimal area surfaces
which correspond to the ground state,
we restrict ourselves to the Y-functions
regular at finite $\theta$.

Taking these into account,
one can write down a set of integral equations
which fully determine the form of Y-functions.
By introducing the following notations
\eqb\label{epsilonandY}
\epsilon(\theta)=\log Y_1(\theta+i\varphi),\qquad
\tepsilon(\theta)=\log Y_2(\theta+i\varphi),
\eqe
the integral equations can be written in the form
of TBA equations
\eqb
\label{TBA1}
\epsilon\Eqn{=}2|Z|\cosh\theta
  +\cK_2\ast\log\bigl(1+e^{-\tepsilon}\bigr)
  +\cK_1\ast\log\bigl(1+\mu e^{-\epsilon}\bigr)
              \bigl(1+\mu^{-1}e^{-\epsilon}\bigr),\\
\label{TBA2}
\tepsilon\Eqn{=}2\sqrt{2}|Z|\cosh\theta
  +2\cK_1\ast\log\bigl(1+e^{-\tepsilon}\bigr)
  +\cK_2\ast\log\bigl(1+\mu e^{-\epsilon}\bigr)
              \bigl(1+\mu^{-1}e^{-\epsilon}\bigr),\qquad
\eqe
where
\eqb\label{kernels}
\cK_1(\theta)=\frac{1}{2\pi\cosh\theta},\qquad
\cK_2(\theta)=\frac{\sqrt{2}\cosh\theta}{\pi\cosh 2\theta},\qquad
\eqe
and
$f\ast g=\int_{-\infty}^\infty
 d\theta' f(\theta-\theta')g(\theta')$.
Note that these equations are valid
in the strip region $-\pi/4<{\rm Im\,}\theta<\pi/4$.
From these equations it is clear that
$\epsilon(\theta), \tepsilon(\theta)$ are even, real functions
\eqb
\epsilon(-\theta)\Eqn{=}\epsilon(\theta),\qquad
\tepsilon(-\theta)=\tepsilon(\theta),\\
\overline{\epsilon(\theta)}\Eqn{=}\epsilon(\bar\theta),\qquad
\overline{\tepsilon(\theta)}=\tepsilon(\bar\theta).
\eqe
In terms of Y-functions, these properties are expressed as
\eqb
\label{Yevenness}
Y_a(-\theta+i\varphi)\Eqn{=}Y_a(\theta+i\varphi)
\eqe
and (\ref{Yreality}).

Historically, the above TBA equations were first constructed
for the $\bbZ_4$-symmetric integrable model.
Indeed, the integral kernels can be obtained as
derivatives of logarithm of the S-matrix elements
in the $\bbZ_N$ model \cite{Koberle:1979sg} with $N=4$.

By using (\ref{Yspecial}),
the cross-ratios (\ref{crossratios})
are expressed in terms of $\epsilon(\theta),\tepsilon(\theta)$.
If one wants to keep $\epsilon,\tepsilon$
evaluated within the above strip region,
the appropriate choice of formulas are
\eqb\label{bU-epsilon1}
b_{k+1}\Eqn{=}
  \exp\biggl[{\epsilon\biggl(\frac{k\pi i}{2}-i\varphi\biggr)}\biggr],
\qquad
U_{k-1}=1+
\exp\biggl[
  {\tepsilon\biggl(\frac{(2k-1)\pi i}{4}-i\varphi\biggr)}\biggr]
\eqe
for
$(2k-1)\pi/4<\varphi<k\pi/2$, and
\eqb\label{bU-epsilon2}
b_{k+1}\Eqn{=}
  \exp\biggl[{\epsilon\biggl(\frac{k\pi i}{2}-i\varphi\biggr)}\biggr],
\qquad
U_{k}=1+
\exp\biggl[
  {\tepsilon\biggl(\frac{(2k+1)\pi i}{4}-i\varphi\biggr)}\biggr]
\eqe
for
$k\pi/2<\varphi<(2k+1)\pi/4$,
where $k=1,2,3$ mod $3$.
The other cross-ratios are obtained by solving
the relations (\ref{crossratios}) and (\ref{b_constraint}).

\begin{figure}[htbp]
\begin{center}
\resizebox{75mm}{!}{\includegraphics{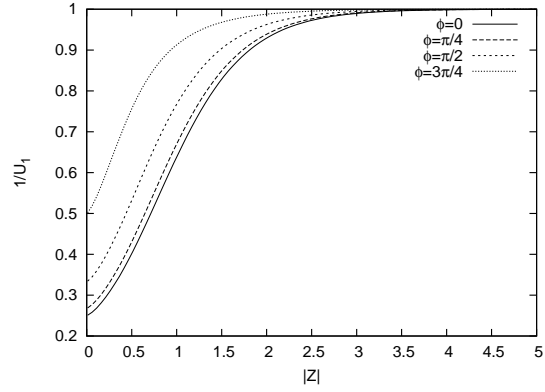}}\\
(a)
\end{center}
\begin{tabular}{@{}c@{}c@{}}
\resizebox{75mm}{!}{\includegraphics{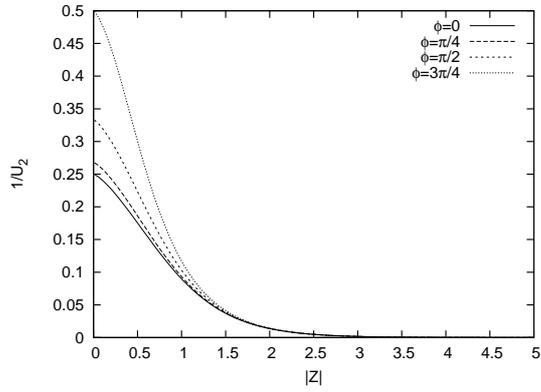}}
&
\resizebox{75mm}{!}{\includegraphics{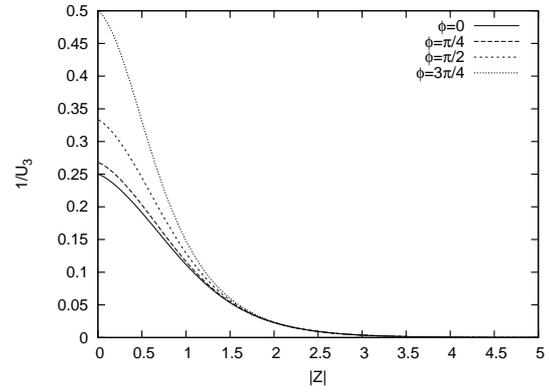}}
\\
(b)&(c)
\end{tabular}
\caption{(a) The $|Z|$-dependence of the function $1/U_1$ for
different values of $\phi$ at $\varphi=-{\pi/48}$.
(b) $1/U_2$. (c) $1/U_3$.\label{fig:z-u_i}}
\end{figure}

Solving the TBA equations numerically, we can calculate the cross-ratios 
as functions of $|Z|$, $\varphi$ and $\phi$. In Figure \ref{fig:z-u_i}, 
we plot $1/U_k$ 
for various $\phi$ at fixed $\varphi=-\pi/48$ as an example.\footnote{
$\varphi=-\pi/48$ is merely a generic value and does not have
particular meaning.}

%

\subsection{Area and remainder function}
As explained in \cite{Alday:2009dv,Alday:2010vh} in detail,
the area of general $n$-cusp solutions with
$n\ne 4k\ (k\in\bbZ_{>0})$
can be written as\footnote{
For $n=4k\ (k\in\bbZ_{>0})$
one needs extra terms \cite{Alday:2010vh,Yang:2010az}.
}
\eqb
A
=A_{\mbox{\scriptsize div}}+A_{\mbox{\scriptsize BDS-like}}
+A_{\mbox{\scriptsize periods}}+A_{\mbox{\scriptsize free}}
+ \mbox{const.}
\eqe
$A_{\mbox{\scriptsize div}}$ and
$A_{\mbox{\scriptsize BDS-like}}$
are contributions from the region inside
the AdS radial coordinate cut-off
and can be evaluated for general $n$.
$A_{\mbox{\scriptsize periods}}$ 
is given by a period integral over the
Riemann surface of the $w$-coordinate
defined by $dw=P(z)^{1/4}dz$.
$A_{\mbox{\scriptsize free}}$
is shown to be identified with (minus)
the free energy of the TBA system.

On the other hand, there is known an all-order
ansatz for the MHV gluon scattering amplitudes
proposed by Bern, Dixon and Smirnov \cite{Bern:2005iz}.
It was shown (assuming the dual conformal symmetry) that
BDS ansatz is correct for $n=4,5$,
but deviates from the string theory computation
for $n\ge 6$ \cite{Alday:2007he}.
The function which complements the ansatz
to produce the full gluon scattering amplitude
is called the remainder function $R$.
In the strong coupling limit, scalar magnitude
of gluon scattering amplitudes
is then expressed as
\eqb
-A=-A_{\mbox{\scriptsize div}}-A_{\mbox{\scriptsize BDS}}+R.
\eqe
$A_{\mbox{\scriptsize div}}$ is identical
to that in the string theory computation,
with appropriate identification of the cut-off parameters.
$A_{\mbox{\scriptsize BDS}}$ can be computed from
one-loop perturbation and is known for general $n$.

Taken altogether, the remainder function in the strong coupling limit
is expressed as
\eqb
R\Eqn{=}\label{eq:R}
  A_{\mbox{\scriptsize BDS}}
-A_{\mbox{\scriptsize BDS-like}}
-A_{\mbox{\scriptsize periods}}
-A_{\mbox{\scriptsize free}}+\mbox{const.}\nn\\
\Eqn{=}R_1
-A_{\mbox{\scriptsize periods}}
-A_{\mbox{\scriptsize free}}+\mbox{const.},
\eqe
where
\eqb
R_1\Eqn{\equiv}
  A_{\mbox{\scriptsize BDS}}-A_{\mbox{\scriptsize BDS-like}}.
\eqe
Below we drop the constant term in (\ref{eq:R}).

\begin{figure}[htbp]
\begin{center}
\begin{tabular}{@{}c@{}c@{}}
\resizebox{75mm}{!}{\includegraphics{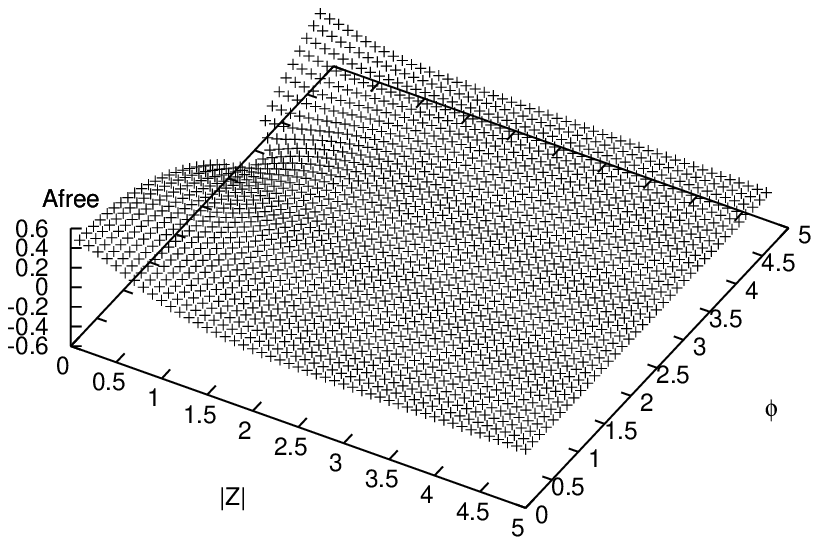}}
&
\resizebox{75mm}{!}{\includegraphics{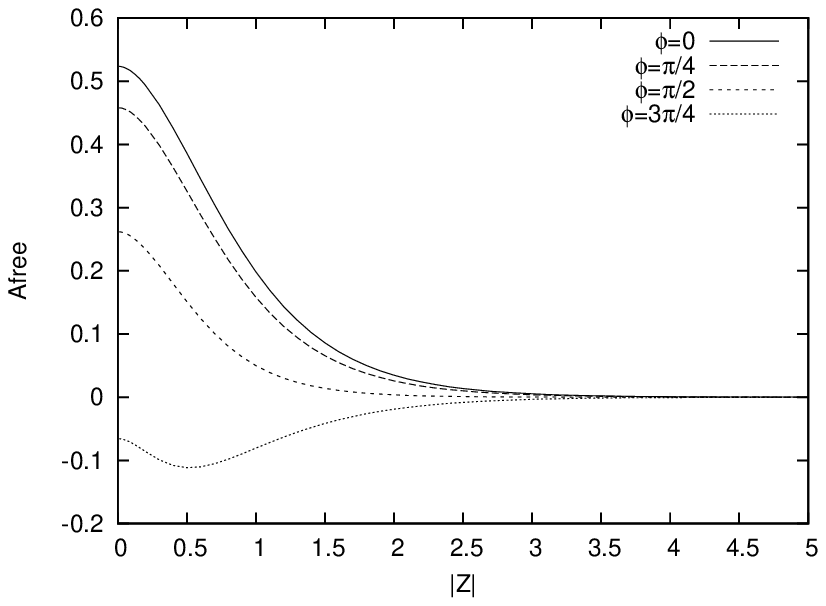}}
\\
(a) & (b)
\\
\end{tabular}
\caption{(a) 3D plots of  
free energy $A_{\mbox{\scriptsize free}}$ 
as a function of $|Z|$ and $\phi$
 ($0\le |Z|\le 5,\ 0\le \phi\le 3\pi/2$)
at $\varphi=-\pi/48$.
(b) $|Z|$-$A_{\mbox{\scriptsize free}}$ graphs for various values of 
$\phi$ at $\varphi=-\pi/48$. \label{fig:z_af}}
\end{center}
\end{figure}

\begin{figure}[htbp]
\begin{center}
\begin{tabular}{@{}c@{}c@{}}
\resizebox{70mm}{!}{\includegraphics{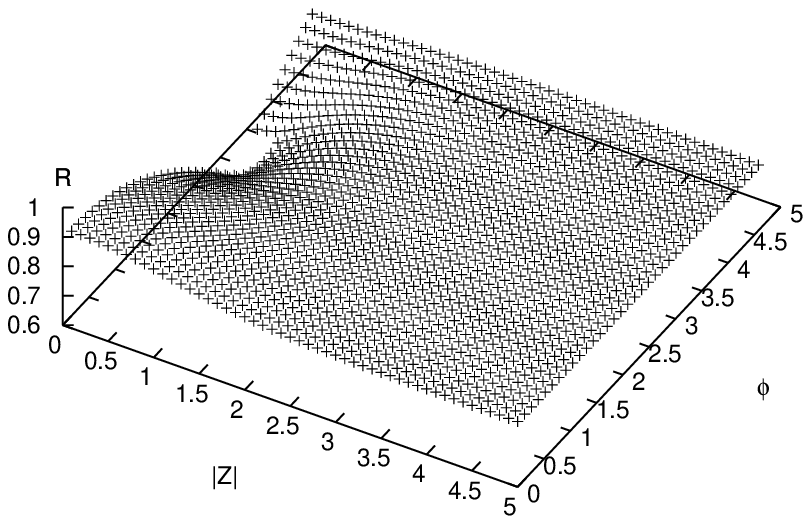}}
&
\resizebox{75mm}{!}{\includegraphics{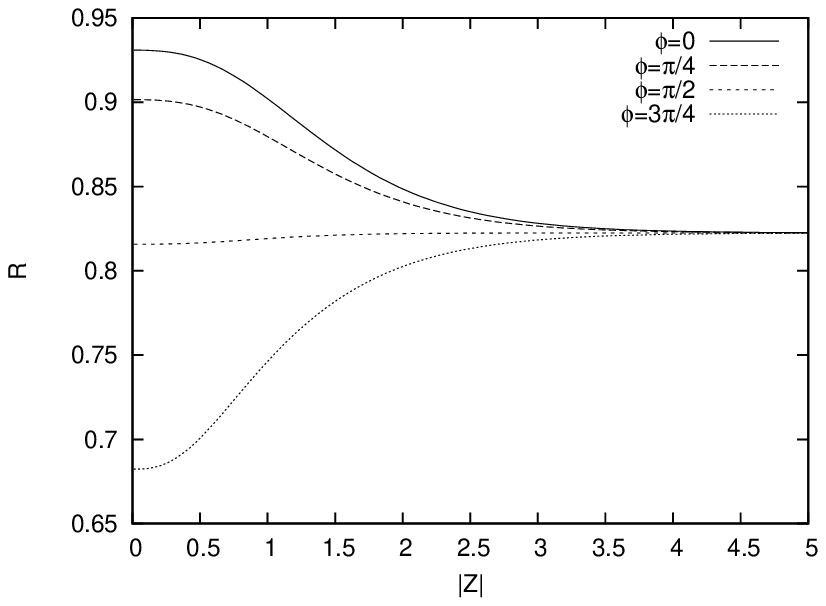}}
\\
(a) & (b)\\
\end{tabular}
\caption{(a) 3D plots of numerical data of 
remainder function $R$ as a function of $|Z|$ and $\phi$ at
 $\varphi=-{\pi/48}$.
(b) $|Z|$-$R$ graphs for various values of $\phi$ at
$\varphi=-{\pi/48}$.\label{fig:z_r}
}
\end{center}
\end{figure}

In the case of $n=6$, terms in (\ref{eq:R}) are
explicitly given by\footnote{
In \cite{Alday:2009dv}
$R_1$ is expressed as
$R_1=\sum_{k=1}^3\left(\frac{1}{8}\log^2u_k
  +\frac{1}{4}\dilog(1-u_k)\right)$
with $u_k=U_k^{-1}$.
This can be rewritten as in (\ref{R1inU})
for $u_k\notin(-\infty,0)$ by using a dilogarithm identity.}
\eqb
R_1
\label{R1inU}
\Eqn{=}-\frac{1}{4}\sum_{k=1}^3 \dilog\left(1 - U_k\right),\\
A_{\mbox{\scriptsize periods}}\Eqn{=}|Z|^2,\\
\label{eq:F}
-F=A_{\mbox{\scriptsize free}}
\Eqn{=}\frac{1}{2\pi}\int_{-\infty}^\infty d\theta
  \biggl(
  2|Z|\cosh\theta\log\bigl(1+\mu e^{-\epsilon(\theta)}\bigr)
  \bigl(1+\mu^{-1}e^{-\epsilon(\theta)}\bigr)\nn\\
\Eqn{}\hspace{5em}
  {}+2\sqrt{2}|Z|\cosh\theta\log\bigl(1+e^{-\tepsilon(\theta)}\bigr)
  \biggr).
\eqe
Figure \ref{fig:z_af} and \ref{fig:z_r} show numerical results 
of the free energy and the remainder 
function at $\varphi=-\pi/48$ respectively, where the value of $\varphi$ 
is the same as in Figure \ref{fig:z-u_i}. From Figure
\ref{fig:z-u_i} and \ref{fig:z_r}(b), we can read off the value of the 
remainder functions as the functions of the cross-ratios $U_k$.

%

%
%

Let us comment on choice of variables.
General $n$-point gluon scattering amplitudes have
$3n-15$ real moduli degrees of freedom.
Correspondingly in the present case, the remainder function $R$
for six-point amplitudes is a function of
three independent real parameters.
One can express $R$, either as a function of
the momentum cross-ratios $U_k$
or as a function of the moduli parameters $|Z|,\varphi,\phi$.
The former choice respects the point of view of the four-dimensional
${\cal N}=4$ gauge theory while the latter fit well with
the two-dimensional description.
In the following sections we mainly adopt the latter picture
and study functional properties of $R$
as well as its constituents $A_{\mbox{\scriptsize free}}$
and $R_1$, in particular in the 
two extreme regions $|Z|\ll 1$ and $|Z|\gg 1$.
The former picture in terms of the cross-ratios is 
also discussed.

\section{Free energy around CFT point}

The TBA equations describing the six-point gluon scattering  
amplitudes (\ref{TBA1})--(\ref{TBA2})
result from minimizing the free energy 
of the two-dimensional $\bbZ_4$-symmetric
integrable field theory \cite{KlaMe},
which is  obtained by deforming the $\bbZ_4$ parafermion
CFT \cite{parafermions}
by the first energy operator \cite{Tsvelick88,Fateev:1990bf}. 
The model contains three particles with mass $m, \sqrt{2}m$ and $m$,
respectively. The third one is the anti-particle of the first. 
The free energy of the model on a space
of length $L \gg 1$ with temperature $1/R$ 
gives the ground state energy of the model on a space 
of length $R$ \cite{Zamolodchikov:1989cf}. 
The TBA equations (\ref{TBA1})--(\ref{TBA2}) tell us that
the scale $R$ is related to $|Z|$ as $mR = 2 |Z|$.
In \cite{Alday:2009dv}, the free energy and the regularized area 
in the CFT limit  $mR \to 0$ are analyzed. 
In this limit, the cross-ratios take the values on the locus 
$U_1=U_2=U_3$.
In this section,
we discuss the corrections to the free energy around  the CFT point.

We begin by noting that the action of the integrable model takes
the form
\eqb\label{action}
S = S_{\rm PF} + \lambda \int d^{2}x \, \varep (x).
\eqe
Here, $S_{\rm PF}$ is the action of the $\bbZ_4$ parafermion CFT,
and $\varep $ is the canonically normalized first energy operator 
with  conformal dimension $D_\varep = \bar{D}_\varep= 1/3$.
The coupling constant $\lambda > 0$ is related to the mass  $m$
as \cite{Fateev:1993av}
\eqb\label{lambdam}
   (2\pi \lambda)^2 = 
   \Bigl[ m \sqrt{\pi} \, 
   \gamma\Bigl(\frac{3}{4}\Bigr)\Bigr]^{\frac{8}{3}}
   \gamma\Bigl(\frac{1}{6}\Bigr),
\eqe
where $\gamma(x) \equiv \Gamma(x)/\Gamma(1-x) $.

The free energy around the CFT point is then
given by the ground state energy $E(R)$ of the  perturbed 
CFT on a cylinder of circumference $R$ with small $\lambda $
in (\ref{action}) \cite{Zamolodchikov:1989cf}:\footnote{
We have rescaled the free energy as $R^2F/L \to F$ in accord with
(\ref{eq:F}).}
\eqb\label{FR}
   F =  R E(R)  - R^2 B(\lambda) ,
\eqe
where $E(R)$ is expanded 
by the connected CFT correlators as
\eqb\label{E0perturbed}
   E(R) \Eqn{=} E_0 - R \sum_{n=1}^{\infty} \frac{(-\lambda)^{n}}{n!} 
     \Bigl( \frac{2\pi}{R}\Bigr)^{2(D_{\varep}-1)n +2}  \\
   &&\, \times  \int \Big\bra V(\infty) \, \varep(z_{n},\zbar_{n}) 
   \cdots \varep(z_{1},\zbar_{1}) \, V(0) \Big\ket_{\rm connected} 
   \prod_{i=2}^{n} (z_{i}\zbar_{i})^{D_{\varep}-1} dz^{2}_{i} .\nn
\eqe
In the above, $E_0$ is the unperturbed ground state energy, 
$V$ is the operator corresponding to
the vacuum,  and the correlators are evaluated on the complex plane
after a conformal transformation from the cylinder. 
We have also set $z_1 = \zbar_1 =1$.
Substituting  $\lambda$ in (\ref{lambdam}), one finds that
the ground state energy has an expansion in $(mR)^{\frac{4}{3}n}$.
Due to the $\bbZ_2$ symmetry $\varep \to - \varep$, only the
terms with even $n$ remain in the expansion \cite{KlaMe}.

The second term in (\ref{FR}) subtracts the bulk contribution to 
$E(R)$, so that the free energy per unit length vanishes at
zero temperature $ R \to \infty$ as implied by (\ref{eq:F})
and the TBA equations. This term is evaluated as \cite{KlaMe} 
\eqb\label{bulkterm}
   B(\lambda) = -\frac{1}{4}m^2.
\eqe
Although the derivation in \cite{KlaMe} is given for $\mu =1$,
the asymptotics needed there may  hold  also  for $\mu \neq 1$ to 
give the same result. We will confirm that (\ref{bulkterm}) is 
in agreement with the numerical results for both $\mu =1$ and
$\mu \neq 1$.

Without the chemical potential, namely, for $\mu =1$,
the vacuum operator $V$ is the identity, and thus the expansion
is straightforward. To proceed in the case with general $\mu$, 
we recall the connection between the $\bbZ_4$ parafermion model
and the spin-$\half$ XXZ (XXZ${}_{1/2}$) model,
the Hamiltonian of which is
\eqb\label{xxz}
H_{XXZ} = \sum_{j=1}^{N} \Bigl[ S_{j}^{x}S_{j+1}^{x}
+ S_{j}^{y}S_{j+1}^{y}
+ \cos \alpha \,  S_{j}^{z}S_{j+1}^{z}\Bigr].
\eqe
In the continuum limit, the spin variables are represented by a
free boson $\Phi(\tau,\sigma)$ \cite{Affleck:1988zj}:
\eqb\label{Sbosonize}
   S_{j}^{z} &\sim& \frac{1}{2\pi r}\del_{\sigma} \Phi
   + (-1)^{j} c_1 \cos (\Phi/r), \nn \\
   S_{j}^{-} &\sim& e^{\frac{i}{2}r \tilde{\Phi}} 
   \Bigl[ c_2 \cos (\Phi/r) + (-1)^{j} c_3 \Bigr],
\eqe
where $S^\pm_j \equiv S_{j}^{x} \pm i S_{j}^{y} $, 
$c_a$ are constants, $\tilde{\Phi}$ is the dual boson,
and $(\tau, \sigma)$ are the two-dimensional coordinates. 
$r$ is the compactification
radius, {\it i.e.}, $\Phi \sim \Phi + 2 \pi r$,
and is related to the coupling 
constant 
by $r = \sqrt{2(1-\frac{\alpha}{\pi})}$ \cite{Luther:1975wr}
in the unit where the selfdual radius is $r_{\rm sd}=\sqrt{2}$.
Since the $\bbZ_4$ parafermion theory is described by a free boson
compactified on the $S^1/\bbZ_2$
orbifold with $r_{\rm PF} = \sqrt{2/3}$ \cite{Yang:1987wk}, 
the continuum limit of the XXZ${}_{1/2}$ model
with $\alpha_{\rm PF} = \frac{2}{3} \pi$ gives 
the $\bbZ_4$ parafermions.

In terms of the XXZ${}_{1/2}$ model, the chemical potential 
$\mu = e^{i\phi}$ in the TBA equations is understood as 
the twist parameter of the boundary conditions\cite{Suzuki,KBP,Fendley},
\eqb\label{twist}
   S^{z}_{N+1} = S_{1}^{z}, \quad 
   S_{N+1}^{\pm} = e^{\pm \frac{2}{3}
   i \phi} S_{1}^{\pm}.
\eqe
The bosonization formula (\ref{Sbosonize}) means that these are    
equivalent to the winding condition on the dual boson 
\cite{Nassar:1997yz,Alday:2010vh},
$\tilde{\Phi}(\tau, \sigma+R)\sim \tilde{\Phi}(\tau,\sigma)
 - \frac{4}{3r} \phi$,
where $R$ is the length of the space.
For the original boson $\Phi$, this induces the shift of the momenta,
\eqb
\frac{n}{R} \to \frac{1}{R}(n-\frac{\phi}{3\pi r}) \quad (n \in \bbZ).
\eqe
One can check that  the ground state energy
is also changed to (see also \cite{Hamer:1987ei}) 
\eqb
E_\Phi= -\frac{\pi}{6R}
  \Bigl( 1-\frac{2\phi^{2}}{3\pi(\pi -\alpha)}\Bigr), 
\eqe
due to the momentum shift, and that  the TBA result
in \cite{Alday:2009dv}
is indeed reproduced for $\alpha= \alpha_{\rm PF}$:
\eqb
  E_0=  E_\Phi \big\vert_{\alpha=\alpha_{\rm PF}} 
  = -\frac{\pi}{6R} \Bigl( 1-\frac{2\phi^{2}}{\pi^2}\Bigr).
\eqe

The above discussion shows that the vacuum operator $V$
for general $\mu$ is identified with the momentum shift operator,
\eqb
V = e^{-i\frac{\phi}{3\pi r} \Phi}
  = e^{-i\sqrt{\frac{1}{6}}\frac{\phi}{\pi}\Phi},
\eqe
when the parafermions are bosonized.
The energy operator $\varep$ is also given simply by the vertex 
operators of $\Phi$, 
since it is in the untwisted sector of the $S^1/\bbZ_2$
orbifold compactification. 
The conformal dimension then determines its bosonized form,
\eqb
  \varep = a_+ \, e^{i \sqrt{\frac{2}{3}} \Phi}
        +  a_-e^{-i \sqrt{\frac{2}{3}} \Phi},
\eqe
where $a_\pm$ are some coefficients including cocycle factors.
Their explicit forms can be found in \cite{Argyres:1992mw}
but, here, we only note  that the products $a_\pm  a_\mp $
are  c-numbers.

\begin{figure}[htbp]
\begin{center}
\begin{tabular}{@{}c@{}c@{}}
\resizebox{75mm}{!}{\includegraphics{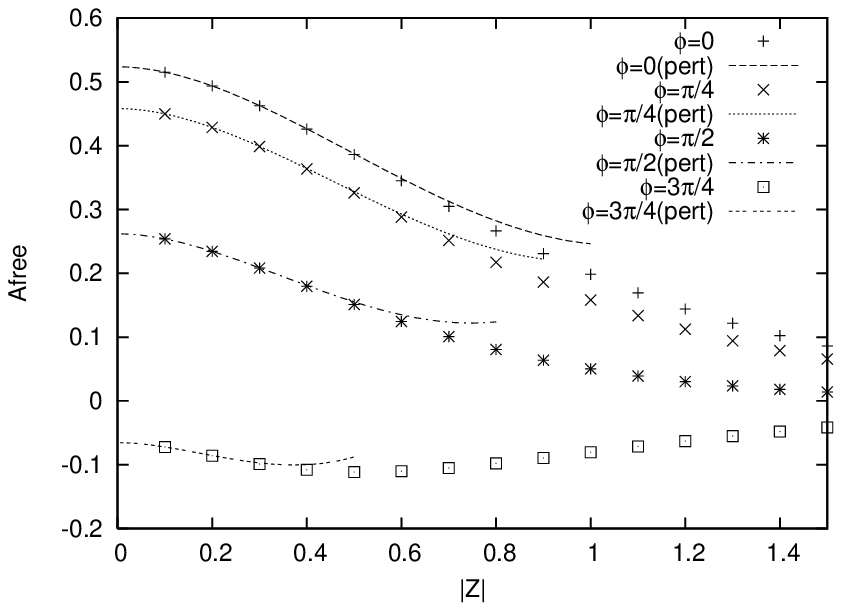}}
&
\resizebox{75mm}{!}{\includegraphics{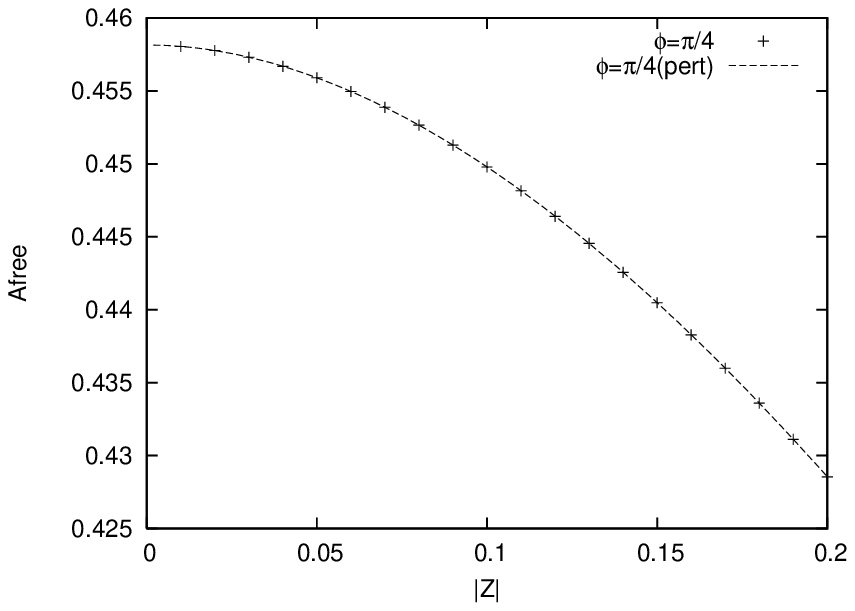}}
\\
(a) & (b)\\
\end{tabular}
\caption{(a) Plots of  $(|Z|,A_{\mbox{\scriptsize free}})$ 
for various values of $\phi$ at $\varphi=-{\pi/48}$.
(b) Plots of $(|Z|,A_{\mbox{\scriptsize free}})$
for $0\leq |Z|\leq 0.2$ at $\phi=\pi/4,\ \varphi=-\pi/48$.
\label{fig:Afree-Z}}
\end{center}
\end{figure}

\begin{figure}[htbp]
\begin{center}
\resizebox{90mm}{!}{\includegraphics{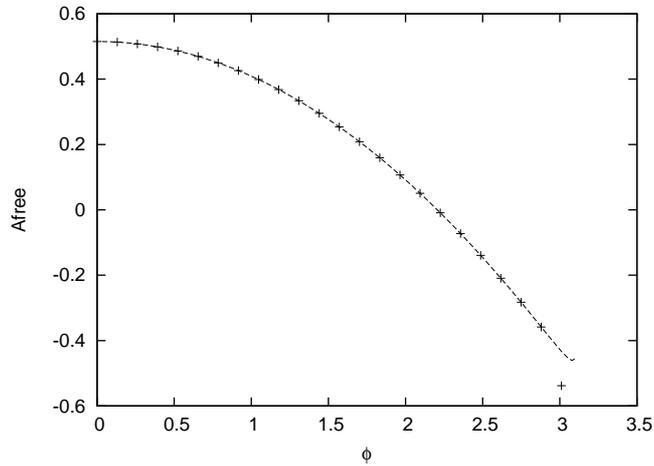}}
\end{center}
\caption{Plots of $A_{\mbox{\scriptsize free}}$
for $0\leq \phi\leq \pi$
 at $|Z|=0.1$ and $\varphi=-{\pi/48}$.
The dashed line represents the graph
calculated by the formula (\ref{eq:free316}).
\label{fig:Afree-phi}}
\end{figure}

Now, we are ready to evaluate the expansion of the ground state energy
$E(R)$. The first non-trivial term in (\ref{E0perturbed}) comes from 
the two-point function.
Since $\varep$ is canonically normalized as
$\langle \, \varep(z) \varep (0) \, \rangle \vert_{\phi=0} =
|z|^{-4/3 }$,  one has  for general $\phi$
\eqb\label{ee}
\Big\langle \varep (z_{2}) \varep(z_1) \Big\rangle_{\phi}
&=& \Big\langle e^{i\sqrt{\frac{1}{6}}\frac{\phi}{\pi}\Phi} (\infty)
  e^{i\sqrt{\frac{2}{3}}
     \Phi}(z_{2}) e^{-i\sqrt{\frac{2}{3}} \Phi}(z_{1}) 
 e^{-i\sqrt{\frac{1}{6}}\frac{\phi}{\pi}\Phi} (0)  \Big\rangle \nn\\
 &=& |1-z_2|^{- \frac{4}{3}}|z_2|^{ \frac{2}{3\pi}\phi},
\eqe
with $z_{1} =1$.
The integral of this term
is evaluated  in (\ref{E0perturbed}) 
by the formula,
\eqb
   \int d^2u \, |u|^{2a} |1-u|^{2b} 
   = \pi \gamma(1+a) \gamma(1+b)\gamma(-1-a-b).
\eqe
Collecting the results so far, we  find the free energy near the CFT
point to be 
\eqb
 F = RE_0 +|Z|^2 - C_{\frac{8}{3}} \,
 \gamma\Bigl(\frac{1}{3} +\frac{\phi}{3\pi}\Bigr) 
 \gamma\Bigl(\frac{1}{3} -\frac{\phi}{3\pi}\Bigr) \, |Z|^{\frac{8}{3}}
 + {\cal O}(|Z|^{\frac{16}{3}}),
\label{eq:free316}
\eqe
where 
\eqb
  C_{\frac{8}{3}} =  \frac{\pi}{2} 
   \Bigl[\frac{1}{\sqrt{\pi}} \gamma
         \Bigl(\frac{3}{4}\Bigr)\Bigr]^{\frac{8}{3}}
      \gamma\Bigl(\frac{1}{6}\Bigr) \gamma\Bigl(\frac{1}{3}\Bigr)  
      \approx 0.18461.
\eqe
At $\phi=\pi$ the coefficient of $|Z|^{\frac{8}{3}}$ diverges,
which implies that the CFT perturbation breaks down there.
The expansions to higher orders are straightforward. 
Figure \ref{fig:Afree-Z} and \ref{fig:Afree-phi} show  
$A_{\rm free} = -F$  from the  above CFT perturbation
and the numerical computation.
The results in (\ref{ee}) and (\ref{eq:free316})
may be continued to imaginary $\phi$.
While real $\phi$ is physical in the context of the XXZ spin-chain,
so is imaginary $\phi$ in the context of thermodynamics.
Note that imaginary $\phi$ corresponds to minimal surfaces
in $(2,2)$ signature \cite{Alday:2009dv}.

%

\section{Remainder function around CFT point}

The remainder function (\ref{eq:R})
essentially consists of two non-trivial functions,
$A_{\mbox{\scriptsize free}}$ and $R_1$.
We studied the structure of $A_{\mbox{\scriptsize free}}$
for $|Z|\ll 1$ in the last section. 
Here let us see how $R_1$ behaves
as a function of $|Z|,\varphi,\phi$ in the region $|Z|\ll 1$.

We know the form of $R_1$ as a function of $U_k$ (\ref{R1inU}).
Using (\ref{Yspecial}), one can express it in terms of Y-functions as
\eqb\label{R1inY}
R_1=-\frac{1}{4}\sum_{k=1}^3
  \dilog\left(-Y_2\left(\frac{(2k+1)\pi i}{4}\right)\right).
\eqe
What we need is the information of
how $|Z|, \varphi,\phi$ enters in the function $Y_2(\theta)$.
From (\ref{epsilonandY})--(\ref{TBA2}) we see that
$\theta$ and $\varphi$ appear
in the Y-functions only through the combination $\theta+i\varphi$.
The Y-functions are periodic in $\theta$ (\ref{Yperiodicity})
and also exhibit $\bbZ_2$ symmetries
(\ref{Yreality}), (\ref{Yevenness}).
Also they are regular everywhere except
at ${\rm Re\,}\theta=\pm\infty$, as we mentioned in section 2.5.
Combining all these properties together, we see that
the Y-functions admit the following Laurent expansion
\cite{Zamolodchikov:1991et}
\eqb\label{YLaurent}
Y_a(\theta)=Y_a^{(0)}(|Z|,\phi)+\sum_{n=1}^\infty
Y_a^{(n)}(|Z|,\phi)(t^n+t^{-n}),\qquad
t=e^{\frac{4}{3}(\theta-i\varphi)},
\eqe
where $Y_a^{(n)}\in\bbR$.
By substituting this into the expression (\ref{R1inY}),
one obtains the following Fourier expansion for $R_1$
\eqb\label{R1Fourier}
R_1=\sum_{m=0}^\infty R_1^{(m)}(|Z|,\phi)\cos 4m\varphi.
\eqe
Note that in deriving this expression, 
there occur cancellations among powers of $t^n$
evaluated at $\theta=(2k+1)\pi i/4,\ k=1,2,3$,
whenever $n$ is not a multiple of 3.
By construction the coefficients $R_1^{(m)}$ are real.
Therefore $R_1$ has shown to be real-valued, even, periodic function
in $\varphi$ with periodicity $\pi/2$.
Our numerical results indeed exhibit this periodicity.

\begin{figure}[htbp]
\begin{center}
\resizebox{90mm}{!}{\includegraphics{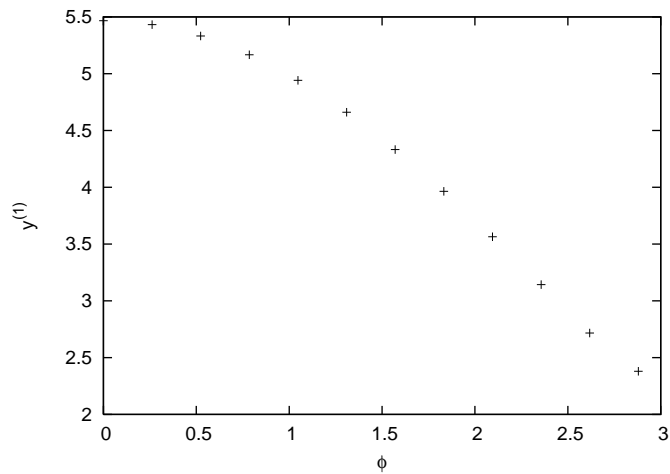}}
\end{center}
\caption{Plots of $y^{(1)}(\phi)$ 
obtained by fitting the numerical data of $Y_2(\theta)$ 
in the region $0.01\leq |Z|\leq 0.2$
at $\varphi=-\pi/48$.
When we fit the numerical data of $y^{(1)}(\phi)$ by the function
$
a+b \phi^2+c \phi^4
$, 
we find that the coefficients are given by
$a=5.47669$, $b=-0.484171$, $c=0.0119471$.
If we fit them by
$
\tilde{a}+\tilde{b}\cos({\phi/3})+\tilde{c}\cos({2\phi/3}),
$
we obtain
$\tilde{a}=1.31367$, $\tilde{b}=2.61136$, $\tilde{c}=1.55402$.
}
\label{fig:u1}
\end{figure}

\begin{figure}[htbp]
\begin{center}
\resizebox{90mm}{!}{\includegraphics{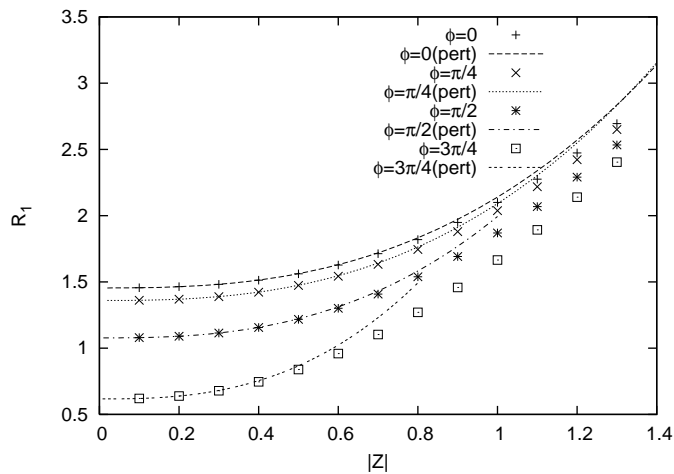}}
\end{center}
\caption{Numerical plots of $|Z|$-$R_1$ vs
perturbative solution of $R_1$ at $\varphi=-\pi/48$.
Here we use  $y^{(1)}(\phi)$ in Figure \ref{fig:u1}.
}
\label{fig:Z-R1}
\end{figure}

Let us now analyze the form of $R_1$ for $|Z|\ll 1$.
It is seen from (\ref{TBA1})--(\ref{TBA2}) that 
the Y-functions grow dramatically at large $\theta$
but do not vary so much in the region
$-\log(1/|Z|)\ll \theta\ll \log(1/|Z|)$.
In other words, Y-functions draw a plateau
over the region, which is wide when $|Z|$ is small.
In order for the Y-functions to be so,
the coefficients $Y_a^{(n)}$ in the expansion (\ref{YLaurent})
have to be sufficiently small and at most
\eqb\label{Ycoeffestim}
Y_a^{(n)}(|Z|,\phi)\sim |Z|^{\frac{4}{3}n}
  +\mbox{(higher order terms)}.
\eqe
The appearance of powers of $|Z|^{4/3}=|U|$ is not totally 
unexpected, as we have already seen in the last section
that the CFT result for $A_{\mbox{\scriptsize free}}$
is obtained in powers of $|Z|^{\frac{8}{3}}$.
This suggests us that the Y-functions,
and thus their descendants $U_k$ and $R_1$,
may well be expanded in powers of $|Z|^{4/3}=|U|$.
Below we assume that this is the case.
Let us then express $Y_2$ as
\eqb
Y_2(\theta)=\sum_{n=0}^\infty
  \tY_2^{(n)}(\varphi+i\theta,\phi)|Z|^{\frac{4}{3}n}.
\eqe
The first coefficient is the value of $Y_2$
evaluated at $Z=0$ and is known \cite{Alday:2009dv} as
\eqb\label{Y_2-0}
\tY_2^{(0)}(\varphi,\phi)
=1+\mu^{2/3}+\mu^{-2/3}=1+2\cos\frac{2\phi}{3}.
\eqe
(\ref{Ycoeffestim}) tells us that only the 
terms with $n=0,1$ in (\ref{YLaurent}) contribute
to the next coefficient $\tY_2^{(1)}$.
Therefore $\tY_2^{(1)}$, as a function of $\varphi$,
is at most the sum of
a term independent of $\varphi$
and a term proportional to 
$t+t^{-1}=2\cos[4(\varphi+i\theta)/3]$.
As a matter of fact,
the former term is not allowed by
the Y-system relations (\ref{Ysystem1}), (\ref{Ysystem2}).
Hence
$\tY_2^{(1)}$ has to be of the form 
\eqb\label{Y_2-1}
\tY_2^{(1)}(\varphi,\phi)
=y^{(1)}(\phi)\cos\frac{4(\varphi+i\theta)}{3}.
\eqe
This form is also confirmed by numerical computations.
At present we do not know the analytic expression for
the function $y^{(1)}(\phi)$, but it can be evaluated
numerically as in Figure \ref{fig:u1}.
Similarly, $\tY_2^{(2)}$ depends on $\varphi$ only through
$\cos [4(\varphi+i\theta)/3]$
and $\cos [8(\varphi+i\theta)/3]$,
but actually we do not need the precise form here.
By using the above data and the Y-system relations
(\ref{Ysystem1})--(\ref{Ysystem2}),
one can determine the behavior of $R_1$ for $|Z|\ll 1$.
If we express $R_1$ in the form
\eqb
R_1=\sum_{n=0}^\infty
    \tR_1^{(n)}(\varphi,\phi)|Z|^{\frac{4}{3}n},
\eqe
the first few coefficients are obtained as
\eqb
\tR_1^{(0)}(\varphi,\phi)\Eqn{=}-\frac{3}{4}\dilog(1-4\beta^2),\\
\tR_1^{(1)}(\varphi,\phi)\Eqn{=}0,\\
\tR_1^{(2)}(\varphi,\phi)\Eqn{=}
  \frac{3(4\beta^2-1+\log(4\beta^2))}{64\beta^2(4\beta^2-1)^2}
  y^{(1)}(\phi)^2,
\eqe
with $\beta=\cos(\phi/3)$.
Note that these three coefficients are $\varphi$-independent.
This is consistent with the argument below (\ref{R1Fourier})
that Fourier modes $\cos(4n\varphi/3)$ with
$n$ not being a multiple of $3$ cancel out in $R_1$.
The $\varphi$-dependence could start appearing at
the order of $|Z|^4$.
The above result agrees with numerical computation with high accuracy.
Figure \ref{fig:Z-R1} shows a comparison
between the above perturbative
approximation with numerical plots.

\begin{figure}[htbp]
\begin{center}
\begin{tabular}{@{}c@{}c@{}}
\resizebox{75mm}{!}{\includegraphics{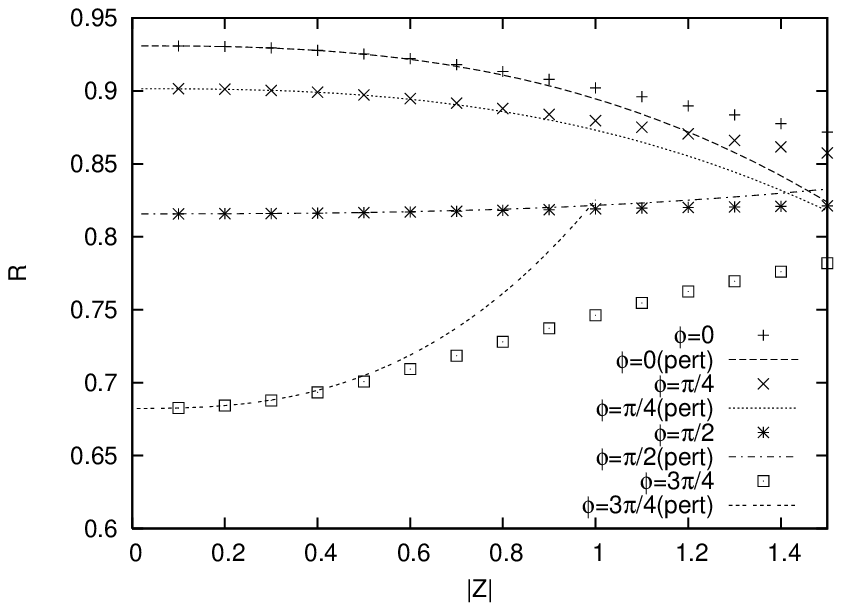}}
&
\resizebox{75mm}{!}{\includegraphics{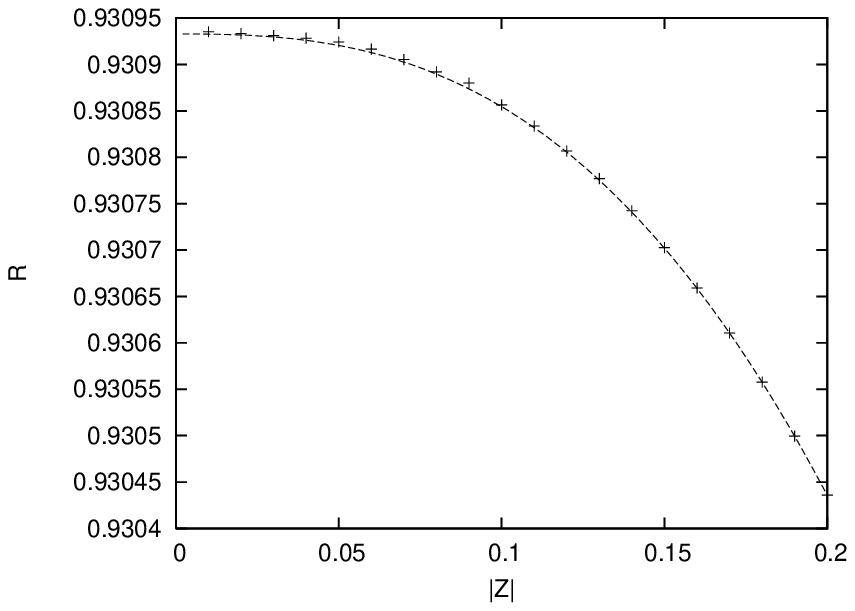}}
\\
(a) & (b)
\end{tabular}
\caption{Numerical plots of $|Z|$-$R$ vs
perturbative solution of $R$ (a) for various values of $\phi$ 
at $\varphi=-\pi/48$ (b)
in the region $0\leq
 |Z|\leq 0.2$ at $\phi=0$ and $\varphi=-\pi/48$.
}
\label{fig:Z-R1-00}
\end{center}
\end{figure}

Collecting the results so far,
the remainder function is expanded around $|Z|=0$ as
\eqb\label{Rexpansion}
R\Eqn{=} -\left[\frac{\pi}{6}\left(1-\frac{2\phi^2}{\pi^2}\right)
  +\frac{3}{4}\dilog(1-4\beta^2)\right]\nn\\
&&+\left[ -
  C_{\frac{8}{3}}\gamma\left(\frac{1}{3}+\frac{\phi}{3\pi}\right)
  \gamma\left(\frac{1}{3}-\frac{\phi}{3\pi}\right)
  +\frac{3(4\beta^2-1+\log(4\beta^2))}{64\beta^2(4\beta^2-1)^2}
  y^{(1)}(\phi)^2
  \right]|Z|^{\frac{8}{3}}\nn\\
&&+{\cal O}\left(|Z|^4\right).
\eqe
Note that $|Z|^2$ terms in $A_{\mbox{\scriptsize periods}}$
and in $A_{\mbox{\scriptsize free}}$ cancel each other.
Figure \ref{fig:Z-R1-00} shows a comparison
between the perturbative
approximation with numerical plots.

One can also express the cross-ratios $U_k$ ($k=1,2,3$)
as
\eqb\label{Ukexpansion}
U_k=4\beta^2
+y^{(1)}(\phi)\left(\cos\frac{4\varphi-(2k+1)\pi}{3}\right)
|Z|^{\frac{4}{3}}
+{\cal O}\left(|Z|^{\frac{8}{3}}\right).
\eqe
These are inverted 
to express $|Z|,\varphi,\phi$ as functions of $U_k$: 
\eqb
\label{phiUk}
   \beta^2 \Eqn{=} \cos^2 \frac{\phi}{3} = \frac{1}{12}(U_1+U_2+U_3),\\
\label{varphiUk}
   \tan \frac{4}{3} \varphi \Eqn{=} 
   \frac{\sqrt{3}(U_2-U_3)}{2U_1 -U_2 -U_3},\\
\label{ZabsUk}
    |Z|^{\frac{4}{3}} \Eqn{=} 
    \frac{-2U_1 +U_2 +U_3}{3 y^{(1)}(\phi) \cos \frac{4}{3} \varphi}.
\eqe
With help of numerical fitting, $ y^{(1)} $ is also 
evaluated, {\it e.g.}, as (see Figure \ref{fig:u1})
\eqb
  y^{(1)}(\phi) \Eqn{\approx} 5.47669  -0.484171\phi^2 
   + 0.0119471\phi^2  \nn \\
   \Eqn{\approx}
  1.31367 + 2.61136 \cos\frac{\phi}{3} +1.55402\cos\frac{2\phi}{3}.
\eqe
Substituting these into (\ref{Rexpansion})
gives an analytic  expansion of $R$ in terms of $U_k$,
which can be directly compared with weak coupling results.
This expansion describes the behavior of $R$
around the locus $U_1=U_2=U_3$.

One can check that the Jacobian of the above change of variables
is proportional to 
$(y^{(1)})^2 |Z|^{\frac{5}{3}} \sin(\frac{2}{3} \phi)$
and the transformation is one-to-one for $|Z| \neq 0$,
$0\le\varphi<3\pi/2$ and $0<\phi<3\pi/2$.
The range of $\varphi$ comes from that of the phase of $U$
as seen in (\ref{ZUrel}),
while the range of $\phi$ corresponds to the region where
minimal surfaces are in
$(1,3)$ and usual $(3,1)$ signatures\cite{Alday:2009dv}.
From (\ref{Ukexpansion}), 
one also finds that  all $(|Z|,\varphi,\phi)$ near the CFT point
correspond to real cross-ratios. 
In the above approximation,
each of $|Z|,\varphi,\phi$ has the following
geometrical meaning in the parameter space $(U_1,U_2,U_3)$:
(\ref{phiUk}) implies that
constant $\phi$ spans a plane
perpendicular to the locus $U_1=U_2=U_3$.
$\phi$ specifies the distance between the plane and the origin.
On this plane, $\varphi$ parametrizes a circle
around the center $U_1=U_2=U_3=4\beta^2$
with a radius of $\sqrt{3/2}\,y^{(1)}(\phi)|Z|^{4/3}$.
Thus, in terms of the cross-ratios,
the weak dependence of $R$ on $\varphi$ for small $|Z|$ observed above
is translated into that on the rotation around the locus $U_1=U_2=U_3$.

%

\section{Remainder function in large mass region}

So far, we have studied the remainder function 
when the mass scale $|Z|$ is small.
In this section, following \cite{Zamolodchikov:1989cf} 
we consider the large mass region,  {\it i.e.},  $|Z| \gg 1$, 
which corresponds to the kinematical region around collinear limits.
In this region, we can solve the TBA equations approximately.
Throughout this section, we consider the case that
$-\pi/4<\varphi<0$.
Other cases can be analyzed similarly
by using (\ref{bU-epsilon1}), (\ref{bU-epsilon2}).

Let us first consider the free energy.
For large $|Z|$, the pseudo-energies,
$\epsilon,\ \tepsilon$,
behave as $\epsilon(\theta) \approx 2|Z| \cosh \theta$
and $\tilde{\epsilon}(\theta) \approx 2\sqrt{2}|Z| \cosh\theta$.
The convolution terms in the TBA equations are suppressed exponentially.
Thus the free energy is evaluated by
\begin{align}
A_{\rm free}\approx \int_{-\infty}^\infty \frac{d\theta}{2\pi}
\left[ \( \mu+\mu^{-1} \) 2|Z| \cosh\theta\, e^{-2|Z| \cosh \theta}
+2\sqrt{2}|Z| \cosh \theta\, e^{-2\sqrt{2}|Z| \cosh \theta} \right] .
\end{align}
One can easily find that the free energy is expressed in terms of
the modified Bessel function of the second kind,
\begin{align}
A_{\rm free}\approx \frac{2|Z|}{\pi}
\left[ \( \mu+\mu^{-1}\) K_1(2|Z|)+\sqrt{2}K_1(2\sqrt{2}|Z|) \right].
\label{eq:Afree-large}
\end{align}
From the asymptotics of the modified Bessel function
we see that $A_{\rm free}$ decays exponentially
as $|Z|$ goes to $\infty$.

Next, let us consider the large $|Z|$ behavior of $R_1$.
The leading correction of $b_1$ is given by
\begin{align}
b_1 &=e^{\epsilon(-i \varphi)} 
\approx e^{2|Z| \cos (\hat{\varphi}-\pi/4)}(1+\delta_1),
\end{align}
where $\hat{\varphi}\equiv \varphi+\pi/4$ and 
\begin{align}
&\delta_1 \equiv\nn\\
& \int_{-\infty}^\infty d\theta
\left[ {\cal K}_2\( i\hat{\varphi}-\frac{\pi i}{4}+\theta\)
  e^{-2\sqrt{2}|Z|\cosh \theta}+\(\mu+\mu^{-1}\) 
       {\cal K}_1\( i\hat{\varphi}-\frac{\pi i}{4}+\theta\)
  e^{-2|Z|\cosh \theta}\right]. \label{eq:delta1}
\end{align}
The kernels are defined in (\ref{kernels}).
Similarly,
\begin{align}
U_2-1=e^{\tilde{\epsilon}(-i\hat{\varphi})}
  \approx e^{2\sqrt{2}|Z|\cos \hat{\varphi}}(1+\delta_2)
\end{align}
with
\begin{align}
\delta_2 &\equiv \int_{-\infty}^\infty d\theta
\left[  2{\cal K}_1\( i\hat{\varphi}+\theta\)
  e^{-2\sqrt{2}|Z|\cosh \theta}+\(\mu+\mu^{-1}\)
  {\cal K}_2\( i\hat{\varphi}+\theta\)
  e^{-2|Z|\cosh \theta}\right]. \label{eq:delta2}
\end{align}
From these and (\ref{crossratios}), (\ref{b_constraint}),
we can compute $b_2$ and $b_3$,
\eqb
b_3 = \frac{U_2}{b_1}
=e^{2|Z|\cos (\hat{\varphi}+\pi/4)}(1+\delta_3),\qquad
b_2 = \frac{b_1+b_3+\mu+\mu^{-1}}{b_1b_3-1},
\eqe
where
\begin{align}
\delta_3 \equiv \frac{\delta_2-\delta_1
 +e^{-2\sqrt{2}|Z|\cos \hat{\varphi}}}{1+\delta_1}.
\end{align}
Thus the other cross-ratios $U_1$ and $U_3$ are given by
\begin{align}
1-U_1&=1-b_2b_3
=-e^{-2\sqrt{2}|Z|\sin \hat{\varphi}}(1+\Delta_1), \\
1-U_3&=1-b_1b_2
=-e^{2\sqrt{2}|Z|\sin \hat{\varphi}}(1+\Delta_3),
\end{align}
with
\begin{align}
\Delta_1 \equiv \frac{(1+\delta_3
  +\mu e^{-2|Z| \cos (\hat{\varphi}+\pi/4)})(1+\delta_3+\mu^{-1}
   e^{-2|Z| \cos (\hat{\varphi}+\pi/4))})}{1+\delta_2}-1, \label{eq:Delta1} \\
\Delta_3 \equiv \frac{(1+\delta_1
  +\mu e^{-2|Z| \cos (\hat{\varphi}-\pi/4)})(1+\delta_1+\mu^{-1}
   e^{-2|Z| \cos (\hat{\varphi}-\pi/4))})}{1+\delta_2}-1. \label{eq:Delta3}
\end{align}
Note that all $\delta_i$ and $\Delta_i$ decay exponentially
as $|Z|$ goes to $\infty$.

\begin{figure}[tb]
  \begin{center}
    \includegraphics[width=10cm]{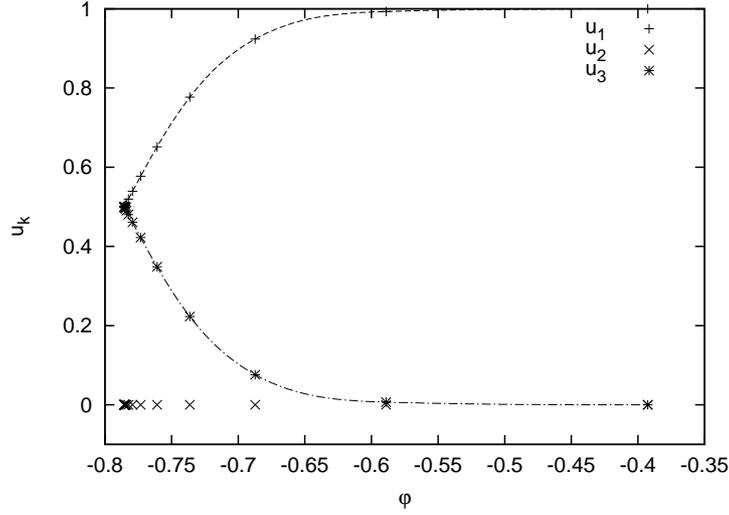}
  \end{center}
  \caption{
We take a large value $|Z|=9.0$ and plot the value of
$u_k (\!{}=\!1/U_k)$
for various $\varphi$ at fixed  $\phi=0$.
Dashed lines correspond to the formula (\ref{eq:col515})
obtained by taking the collinear limit.
}
  \label{fig:psi-u1-u2-u3-00}
\end{figure}

Here let us give a comment on the large $|Z|$ behaviors of
the cross-ratios. When $\hat{\varphi}$ is far from zero,
the cross-ratios
show the asymptotic behavior
$(u_1,u_2,u_3) \to (1,0,0)$
as $|Z| \to \infty$
where $u_k \equiv 1/U_k$ ($k=1,2,3$). 
However if $\hat{\varphi}$ approaches zero, $(u_1,u_2,u_3)$
can reach an arbitrary point on the segment $(1-c,0,c)$ ($0<c<1/2)$.
To see this, we need to take the double scaling limit
$\hat{\varphi} \to +0,|Z| \to \infty$ with
$|Z|\sin \hat{\varphi}=a(>0)$ held fixed.
In this limit
the cross-ratios go to the point
(see Figure \ref{fig:psi-u1-u2-u3-00})
\begin{align}
(u_1,u_2,u_3) \to (1/(1+e^{-2\sqrt{2}a}),0,1/(1+e^{2\sqrt{2}a})).
\label{eq:col515}
\end{align} 
In order to invert the relation between $(|Z|,\varphi,\phi)$
and the cross-ratios for large $|Z|$, one has to evaluate
the integrals in (\ref{eq:delta1}) and (\ref{eq:delta2}) in more detail.

To analyze the large $|Z|$ behavior of $R_1$,
the following asymptotic expansion is useful:
\begin{align}
\Li_2(-e^{x}(1+\delta))
=-\frac{x^2}{2}-\frac{\pi^2}{6}-x \delta +e^{-x}
+{\cal O}(\delta^2,e^{-x} \delta , e^{-2x}) \quad (x \gg 1 \;\;
{\rm and} \;\; \delta \ll 1).
\end{align} 
Therefore the asymptotic behavior of $R_1$ is given by
\eqb
R_1
\Eqn{\approx} |Z|^2+\frac{\pi^2}{12}\nn\\
&&+\frac{1}{4}( 2\sqrt{2}|Z| \cos \hat{\varphi}\;
 \delta_2+2\sqrt{2}|Z|\sin \hat{\varphi}\; \Delta_3
-e^{-2\sqrt{2}|Z|\cos \hat{\varphi}}
+e^{-2\sqrt{2}|Z|\sin\hat{\varphi}}\Delta_1
).\nn\\
\label{eq:R1-large}
\eqe
The first term is divergent in the large $|Z|$ limit,
but this term is canceled by the second term in \eqref{eq:R}.
Combining all the above results,
we finally arrive at the large mass behavior of the remainder function
\begin{align}
\label{eq:R-largeZ}
R&\approx \frac{\pi^2}{12}-\frac{2|Z|}{\pi}
\left[ \( \mu+\mu^{-1}\) K_1(2|Z|)+\sqrt{2}K_1(2\sqrt{2}|Z|) \right]\\
&\quad +\frac{1}{4}( 2\sqrt{2}|Z| \cos \hat{\varphi}\;
 \delta_2+2\sqrt{2}|Z|\sin \hat{\varphi}\; \Delta_3
-e^{-2\sqrt{2}|Z|\cos \hat{\varphi}}
+e^{-2\sqrt{2}|Z|\sin\hat{\varphi}}\Delta_1
).\nn
\end{align}
where $\delta_2$, $\Delta_1$ and $\Delta_3$ are given by
\eqref{eq:delta2}, \eqref{eq:Delta1} and \eqref{eq:Delta3} respectively.
Since the second and the third terms
decay exponentially
to zero in the large mass limit, 
the remainder function approaches the constant $\pi^2/12$.

\begin{figure}[tb]
  \begin{center}
\begin{tabular}{@{}c@{}c@{}}
\resizebox{75mm}{!}{\includegraphics{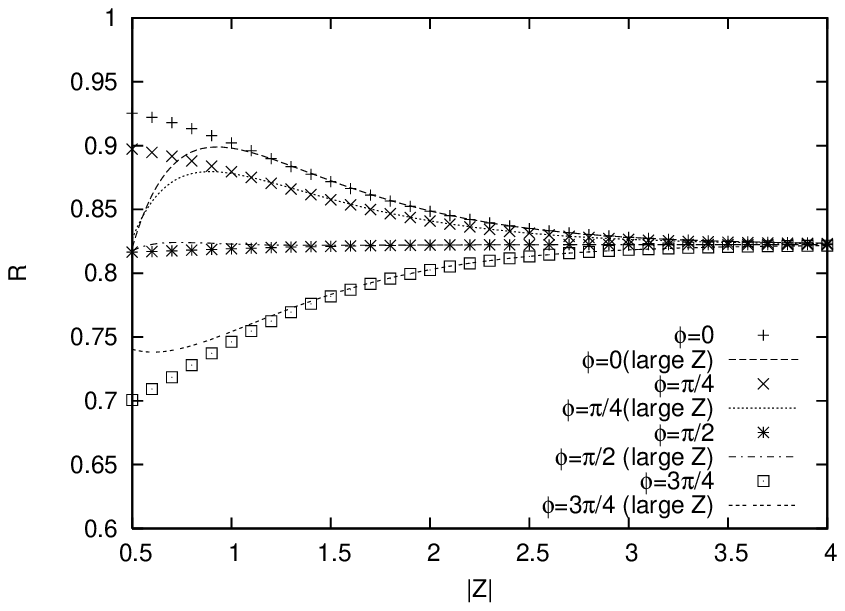}}
&
\resizebox{75mm}{!}{\includegraphics{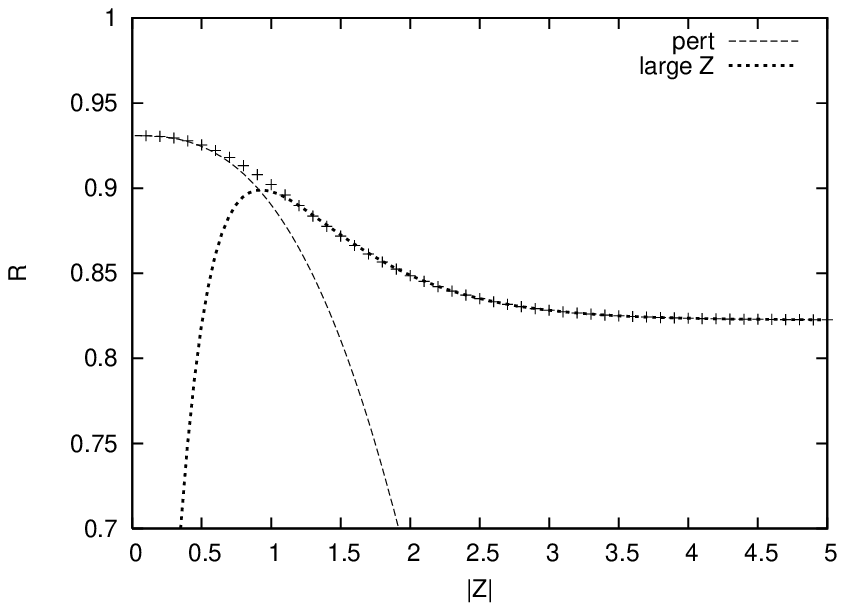}}
\\
(a) & (b)
\end{tabular}
\caption{(a) Large $|Z|$ behavior of the remainder function for
various $\phi$ at $\hat{\varphi}=5\pi/24$.
  Dashed lines show the asymptotic expansion 
obtained by (\ref{eq:Afree-large}) and (\ref{eq:R1-large}). 
  Points show the numerical results.
  \label{fig:Z-R-largeZ}
(b) The $|Z|$-dependence of the remainder function vs 
the small (dashed line) and the large (dotted line) $|Z|$
expansions at $\phi=0$ and $\varphi=-{\pi/48}$.
The two expansions cover the whole region in $|Z|$ except 
a small region around $|Z|=1.0$.}
\end{center}
\end{figure}
%

\begin{figure}[tb]
  \begin{center}
\begin{tabular}{@{}c@{}c@{}}
\resizebox{75mm}{!}{\includegraphics{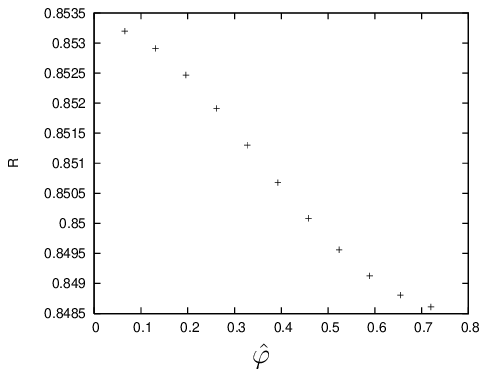}}
&
\resizebox{75mm}{!}{\includegraphics{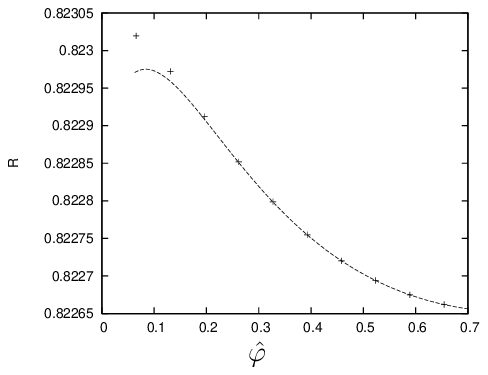}}
\\
(a) & (b)
\end{tabular}
\caption{$\hat{\varphi}$-dependence of the remainder function
(a) at $|Z|=2.0$ and $\phi=0$ and (b) at $|Z|=5.0$ and $\phi=0$.
  The dashed line and points correspond to the asymptotic expansion
  and the numerical results, respectively.
We see that $R$ does not depend very much on $\hat\varphi$.
($R$ varies only up to $0.6$\% at $|Z|=2.0$
and $0.05$\% at $|Z|=5.0$.)}
  \label{fig:psi-R-largeZ}
  \end{center}
\end{figure}

Figure \ref{fig:Z-R-largeZ} shows the behavior of
the remainder function in the large $|Z|$ limit.
For the large $|Z|$, the asymptotic expansion of
the remainder function \eqref{eq:R-largeZ} is in good agreement
with the numerical data.
Figure \ref{fig:psi-R-largeZ} shows the $\hat{\varphi}$-dependence
of the remainder function for the large $|Z|$.
We observe that $\hat\varphi$-dependence is weak.
This can also be seen in the analytic form (\ref{eq:R-largeZ})
where the $\hat\varphi$-dependence is suppressed
exponentially in $|Z|$.

\section{Conclusions}

In this paper we have studied the remainder function of 6-point gluon
scattering amplitudes in ${\cal N}=4$ super Yang--Mills theory by
analyzing the TBA equations perturbatively and numerically.
We have examined perturbative solution near the UV and IR limits and 
found that results are consistent with the numerical results.
The remainder function is made of the free energy of
the $\bbZ_4$-symmetric integrable model 
and the difference between
the BDS part and the BDS-like part.
The free energy near the CFT point can be obtained by the correlation
functions with the chemical potential background.
It is an interesting problem to generalize this result to
the case of $n$-point  amplitudes corresponding to the
generalized parafermions.
The other part of the remainder function $R_1$ is written as the sum 
of the dilogarithm function including Y-functions as arguments.
In the present work, we could not completely
determine its analytic form near the CFT point, but the undermined
function of the chemical potential has been
evaluated by numerical fitting.
It is also an interesting problem to determine 
the series expansions of the Y-functions in order to know analytical
properties of the remainder functions.
In the large mass limit, we have obtained the first order correction to 
the remainder function. It would be useful to apply nonlinear integral 
equation approach \cite{NLIE} to analyze the TBA system further.

Our results provide an analytic
form of the remainder function for the 6-point amplitude at strong
coupling away from the UV(CFT) and the IR(collinear) point. 
Such an analytic form beyond numerical ones will 
be important for further studying  the super Yang--Mills theory
at strong coupling.
In particular,  together with the recent results of  the analytic form
at weak coupling \cite{Smir,Zhang:2010tr}, which is still under
active investigation,
our strong coupling result will give a clue to understand the
scattering amplitude to all order
in the context of the AdS/CFT correspondence.
In this regard, comparisons with  the
perturbative (both analytic and numerical) computations 
\cite{Dr2,nume,Smir,Zhang:2010tr} would be of interest.
One can expect that the physical picture of the amplitude to all order
will emerge through further investigations  both at weak and strong
coupling.
As shown in Figure \ref{fig:Z-R-largeZ}, we also find that 
simple first order expansions away from 
the UV and IR points give a good approximation to the remainder 
function for all the scale $|Z|$ as long as the expansions are valid.
In addition, our results demonstrate that 
the identification of the two-dimensional integrable models
underlying the four-dimensional super 
Yang--Mills theory \cite{Alday:2009dv,Hatsuda:2010cc}
is useful, as well as interesting, for actual computations
of the amplitude at strong coupling. The discussion in this paper
may also be generalized to other TBA systems with chemical potential.

The free energy $A_{\mbox{\scriptsize free}}$ is independent
of $\varphi$, the phase of $Z$.
From perturbative and numerical analysis,
we observe that the $\varphi$-dependence of
the remainder function is also weak: it starts appearing
possibly at the order of $|Z|^4$ for small $|Z|$
and at the order of $e^{-c |Z|}$
with $c$ being some positive constant for large $|Z|$.
In terms of the cross-ratios $U_k$, 
the weak dependence for small $|Z|$ is translated into that 
on the rotation around the locus $U_1=U_2=U_3$.
It would be important to further investigate the meaning of the
$\varphi$-dependence to explore quantum corrections to the present
analysis as noted in \cite{Alday:2010vh}.

\vspace{3ex}

\begin{center}
  {\bf Acknowledgments}
\end{center}

We would like to thank J.~Suzuki for useful discussions.
Y.~S. would like to thank J.~Balog for useful discussions and
conversations.
The work of K.~S. and Y.~S. is supported in part by Grant-in-Aid
for Scientific Research from the Japan Ministry of Education, Culture, 
Sports, Science and Technology.

%
%
\def\thebibliography#1{\list
{[\arabic{enumi}]}{\settowidth\labelwidth{[#1]}\leftmargin\labelwidth
  \advance\leftmargin\labelsep
  \usecounter{enumi}}
  \def\newblock{\hskip .11em plus .33em minus .07em}
  \sloppy\clubpenalty4000\widowpenalty4000
  \sfcode`\.=1000\relax}
\let\endthebibliography=\endlist
\vspace{3ex}
\begin{center}
{\bf References}
\end{center}
\par \vspace*{-2ex}

\end{document}